\documentclass[preprint,12pt]{elsarticle}

\usepackage{graphicx}      
\usepackage{amsmath,amssymb,amsfonts}

\usepackage{algorithm,algpseudocode}
\algnewcommand{\Inputs}[1]{%
  \State \textbf{Inputs:}
  \Statex \hspace*{\algorithmicindent}\parbox[t]{.8\linewidth}{\raggedright #1}
}
\algnewcommand{\Initialize}[1]{%
  \State \textbf{Initialize:}
  \Statex \hspace*{\algorithmicindent}\parbox[t]{.8\linewidth}{\raggedright #1}
}

\usepackage{textcomp}
\usepackage[utf8]{inputenc}
\usepackage{color}

\usepackage{lipsum}

\usepackage{amsbsy}
\usepackage{multirow}

\usepackage{subfig}
\usepackage{amsfonts, amsmath, amsthm, amsbsy,amssymb,bm,mathtools} 
\usepackage{graphicx} 		
\usepackage{transparent}	
\newsubfloat{figure}		

\usepackage[dvipsnames]{xcolor}

\usepackage{lineno,hyperref}
\usepackage{mathtools}
\usepackage{verbatim}
\usepackage{comment}

\allowdisplaybreaks[1]
\journal{Journal of Computational Physics}

\begin{document}
\newcommand{\eric}[1]{#1}

\newcommand{\ericc}[1]{\textcolor{ProcessBlue}{--EA: #1}}
\begin{frontmatter}



\title{Identifying Large-Scale Linear Parameter Varying Systems with Dynamic Mode Decomposition Methods}

\author[UFSC]{Jean~Panaioti~Jordanou}
\author[UFSC]{Eduardo Camponogara}
\author[Texas]{Eduardo Gildin}

\address[UFSC]{Department of Automation and Systems Engineering, Federal University of Santa Catarina, Florianópolis, 88040-900, Santa Catarina, Brazil}
\address[Texas]{Harold Vance Department of Petroleum Engineering, Texas A\&M University, College Station, 77843-3116, Texas, USA}         

\begin{abstract}
Linear Parameter Varying (LPV) Systems are a well-established class of nonlinear systems with a rich theory for stability analysis, control, and analytical response finding, among other aspects.
Although there are works on data-driven identification of such systems, the literature is quite scarce regarding the identification of LPV models for large-scale systems.
Since large-scale systems are ubiquitous in practice, this work develops a methodology for the local and global identification of large-scale LPV systems based on nonintrusive reduced-order modeling.
The developed method is coined as DMD-LPV for being inspired in the Dynamic Mode Decomposition (DMD).
To validate the proposed identification method, we identify a system described by a discretized linear diffusion equation, with the diffusion gain defined by a polynomial over a parameter.
The experiments show that the proposed method can easily identify a reduced-order LPV model of a given large-scale system without the need to perform identification in the full-order dimension, and with almost no performance decay over performing a reduction, given that the model structure is well-established.
\end{abstract}


\end{frontmatter}



\section{Introduction}

Linear Parameter Varying (LPV) Systems are a class of nonlinear dynamic models studied in the literature and widely used in applications  \cite{Hoffman2014}. 
In fact, not only are methods for control of LPV systems well established in the literature, but any nonlinear dynamic system can also be represented as the so-called quasi-LPV system.
LPV systems are expressed in terms of a linear relation between the state and input to the state derivative (continuous time) or the next state in sampling time (discrete time).
However, they are nonlinear in the sense that they depend on an extra set of inputs, referred to as parameters.
If the parameters are fully exogenous, the system is called a fully-fledged LPV system. 
However, when the parameters depend on state and inputs (as generally happens when converting a nonlinear system to an LPV system), the LPV system is referred to as a quasi-LPV system.
Works such as \cite{Hoffman2014} show that not only can LPV-type models represent a wide variety of systems, such as airplanes, hypersonic vehicles, satellites, laser printers, drilling rigs, and wind energy generation systems, among many others, but also have very well-defined methods to mathematically design and tune controllers, which is something that nonlinear systems generally lack.
Also, quasi-LPV systems have a mature development in Model Predictive Control (MPC) applications \cite{Morato2024}, where convergence properties are very well defined.
These features prove that there are many advantages in modeling a given system as an LPV.

There are also several methods for identifying LPV systems \cite{LopesdosSantos2011}, which are generally divided into \textit{local} identification \cite{Zhang2017232} and \textit{global} identification approaches \cite{Armanini2018}.
The main difference between them is that the former utilizes data assuming a fixed value for the parameters, while identifying a linear system and then performing interpolation between each system found.
The latter approach uses data from many points, treating the parameters as extra model inputs (feature generators). The local approach is easier to compute but only sometimes feasible, especially in systems that demand quasi-LPV models.
There are some gaps in LPV identification theory, such as how exactly the structure of the scheduling function is defined \cite{LopesdosSantos2011} (in the case of this work, we consider that model structure is known to simplify this issue).
However, the main gap that this work addresses is the lack of approaches in the literature for performing LPV identification of large-scale systems.

Any system is considered large-scale if it has a high number of states and inputs in a way that computing those systems is not trivial, let alone identifying them.
In fact, any system described as a PDE has its discretization presented as a large-scale system since a PDE can be seen as a system with infinite states, and the more refined the discretization, the larger the number of states.
The field that deals with large-scale systems-related problems is called Model Order Reduction (MOR) \cite{dragoslav}.
Model order reduction can be performed intrusively or non-intrusively.
Intrusive MOR involves having a full system and obtaining its reduced form.
An example of intrusive MOR is applying Proper Orthogonal Decomposition (POD) by itself \cite{Chatu2010}, which obtains a linear map between a reduced state space and the full state space of the system, while accounting for the linear map in the formulation.
Non-intrusive MOR involves avoiding computing the system in its full state, since the computer is assumed to be unable to handle the computation.
A large-scale LPV system naturally leads to a combinatorial explosion, which makes computation very challenging. Therefore, to tackle the problem of identifying large-scale LPV systems, we must employ non-intrusive MOR methods.

A well-established non-intrusive MOR method is the Dynamic Mode Decomposition (DMD) \cite{Schmid2022}, which started as a method for identifying reduced-order counterparts as autonomous linear systems.
Since the response of autonomous linear systems is known analytically, it is easy to obtain the approximation of the analytical response of the full-order system through the eigenvalues of the reduced matrix and the so-called dynamic modes, which are approximations of the eigenvectors of the full-order system computed from its reduced-order counterpart.
DMD is also intimately related to the Koopman operator theory, which represents a nonlinear, finite-dimensional system through a linear but infinite-dimensional operator that is a function of observables (functions of the system states).
The linear relation obtained by DMD can be interpreted as a finite approximation of the infinite-dimensional Koopman operator.

Naturally, DMD evolved over time, with tweaks developed for many different applications in mind \cite{Schmid2022}, with the first important one being the Exact DMD, which proposes an alternative way to compute the dynamic modes by using output data information and, as a result, a more precise calculation of the eigenvectors is obtained.
Relevant to the LPV system identification, however, is the so-called DMDc \cite{DMDc,Schmid2022}, where the DMD is reformulated assuming the presence of an exogenous input (be it a disturbance or a control input), which changes the identification procedure by a small margin.
In this work, we see this change as a clear separation between the DMD steps of solving a rank-limited linear least squares (Procrustes Problem \cite{Baddoo2022}) and the POD being performed over a state snapshot matrix to reduce the state space afterward.
In DMD for autonomous systems, the formulation allows the two steps to be performed at the same time, which is why the same data matrix is used to solve both POD and Procrustes.

Another important branch of DMD is the extended DMD (EDMD) \cite{Williams2015,Schmid2022}, where the direct connection between DMD and the infinite-dimensional Koopman operator is exploited since a DMD, considering nonlinear functions of states as features, is seen as an approximation with a linear finite-dimensional operator and a finite number of observables.
However, in EDMD, both the input and output are subject to the mapping to the observables space, and the Procrustes problem is solved in the observable dimension, which is intrinsically quite large.
In other works, such as in \cite{Gosea2021}, where a bilinear model is considered, the nonlinear terms are incremented into the DMD similarly to the control action introduced in DMDc.

As we remark in this work, a large-scale LPV system can suffer from a combinatorial explosion if the number of parameters and states is large.
Hence, this work proposes an efficient method for obtaining reduced-order proxy models from large-scale LPV systems using the DMD theory.
The contributions of this work are as follows:
\begin{itemize}
    \item Since the least square problem associated with large-scale LPV models tends to be very hard to solve, we test the capacity to obtain quality solutions by solving an $r$-rank Procrustes problem instead.
    \item We assess the capacity of performing a DMD identification similar to \cite{Gosea2021} in the context of LPV systems.
    \item For local identification, we explore the possibility of employing the POD transform obtained by the full outputs in each LTI system being identified. As a result, the LTI systems are all identified in a reduced state space, enhancing scalability while being coherent with the nonintrusive model order reduction philosophy.
    \item We study and develop methods to mitigate the combinatorial explosion resulting from the Kronecker product between the number of parameter functions and the number of states.
\end{itemize}

This work is structured as follows:
Section \ref{sec:rel_work} presents a literature review of related works. In contrast, Section \ref{sec:lpv} gives a brief description of LPV systems, discusses standard LPV system identification methods, and describes POD model order reduction for LPV systems.
 Section \ref{sec:dmdc} explains DMDc, a basis for the proposed method.
Section \ref{sec:dmd_lpv} discusses the proposed DMD-LPV method. This method is designed to identify a reduced-order model of Large-Scale LPV systems, offering a practical and efficient solution to a complex problem.
 Section \ref{sec:application} showcases the case studies and experiments performed.
 Section \ref{sec:conclusion} concludes this work.


\section{Related Work} \label{sec:rel_work} 

An example of a recent work performing identification for an LPV model in a given application is \cite{Sheikh2024}, where LASSO and ridge regression are integrated into the LPV identification to obtain a sparse LPV model.
The LPV identification in question is an input-to-output model that was identified globally. The work mentioned considers a lithium-ion battery to be modeled as an LPV system.
It shows that LPV identification is a valid method to tackle some relevant modeling for control and engineering problems.
A work related to large-scale LPVs is \cite{Dehghani2024}, which relies on a large-scale LPV system to design a robust controller for a large power system.
The work considers model order reduction a valid strategy for tackling control problems in large-scale LPV systems.
Also, \cite{Liu2024} develops a strategy for LPV identification with colored noise, working mainly with global identification and employing a three-tank system as a case study.
For a local approach, \cite{Bombois2021} presents a method for optimally selecting the parameter configuration for LTI local systems.
However, none of the aforementioned identification applications tackle the problem of identifying a large-scale LPV system.

In our work, the methods for performing non-intrusive LPV modeling are derived from the DMD theory.
A work that employs similar principles to identify nonlinear systems is \cite{Jiang2024}, which utilizes a variation of DMDc to identify a microgrid system, a very prominent large-scale system application in the form of a nonlinear system.
DMD, by itself, is a powerful tool.
Works such as \cite{Libero2024} employ DMD to perform several dynamic analyses of complex and large-scale systems such as the GRACE satellite.
In principle, DMD-LPV is quite close to the bilinear DMD (biDMD), which was formulated and employed in \cite{Goldschmidt2021} for quantum systems control, as the quantum Hamiltonian has a bilinear structure in its equation, enabling such method to be employed.

The field of combining DMD with LPV systems is still new, so the literature on methodologies for LPV system identification in larger-scale systems is scarce. However, some relevant publications on methods for LPV identification of large-scale systems can be found in the literature.
The first of them is Stacked DMD, which was reviewed on \cite{Huhn2023}.
The method considers only one parameter, applying a local identification approach.
It involves stacking training data of different parameter configurations and performing a single Singular Value Decomposition (SVD).
The review of stacked DMD in \cite{Huhn2023} brings about some drawbacks, such as parameter-independent eigenvalues and the prohibitive computational cost of SVD.
Further, employing such an algorithm in a multi-parameter situation would be complicated.
The method \cite{Huhn2023} proposes identifying different LTI systems and performing Lagrangian interpolation in them while assuming one parameter.
They propose two variants where either the full or the reduced-order system is interpolated, with the interpolation on the reduced system being more successful in the proposed application (parametric PDEs).
The main difference between \cite{Huhn2023} and our approach is that we try to identify the parameter-system relation in the context of DMD instead of performing polynomial interpolation, making assumptions on the scheduling function of the system.

Another work, \cite{Sun2023}, tries identifying a Radial Basis Function Network to relate a single parameter to a set of snapshot matrices, performing DMD to a parameter-dependent matrix function.
The purpose is to assess a plant's dynamic modes at a given operating point. However, no identification of an LPV system per se was performed.

\section{Linear Parameter Varying Systems} \label{sec:lpv}

Linear parameter varying systems are a class of nonlinear systems with input-to-output linearity. However, the model coefficients depend on a different set of inputs, referred to as parameters.

Different forms of expressing an LPV model exist, mainly state-space and input-to-output \cite{lpvs}.
For this work, we consider a LPV system discrete-time of the following state-space form:
    \begin{align} \label{eqn:lpvs}
    \mathbf{x}[k+1] = \mathbf{A}(\Theta[k])\mathbf{x}[k] + \mathbf{B}(\Theta[k])\mathbf{u}[k]
    \end{align}
where $\mathbf{A}(\Theta[k]), \mathbf{B}(\Theta[k])$ are both matrix functions of the parameter vector $\Theta[k]$, $\mathbf{x}[k]$ are the system states, and $\mathbf{u}[k]$ are the inputs.
All variables above are functions of discrete time $k \in \{0,1,2,\hdots,\infty\}$.
When $\Theta[k]$ depends on either the states or the inputs, the LPV system is referred to as a quasi-LPV model \cite{cisneros2018}.
Although this form describes the dynamic portion of an LPV state-space model, the states are treated as direct outputs, which makes the LPV model a first-order input-to-output model.
LPV systems can also be expressed in continuous time, as some examples below may show, but the discussion regarding the properties mentioned in this paragraph remains the same.
Converting continuous-time to discrete-time dynamics is a trivial matter, as many discretization methods are widely known in the literature.

LPV systems are intrinsically related to fully nonlinear systems, represented as follows:
\begin{equation}
    \mathbf{x}[k+1] = \mathbf{f}(\mathbf{x}[k],\mathbf{u}[k])
\end{equation}

We approximate the fully nonlinear system into an LPV representation \cite{lpvs_incremental} by simply employing Taylor's linearization:
\begin{equation}
\left \{
\begin{aligned} \label{eqn:incremental_model}
    \Delta \mathbf{x}[k+1] &= \partial_{\mathbf{x}[k]}\mathbf{f}(\mathbf{x}[k],\mathbf{u}[k])\Delta\mathbf{x}[k] + \partial_{\mathbf{u}[k]}\mathbf{f}(\mathbf{x}[k],\mathbf{u}[k])\Delta \mathbf{u}[k]\\
    \mathbf{x}[k+1] &= \mathbf{x}[k] + \Delta \mathbf{x}[k+1]\\
    \mathbf{u}[k+1] &= \mathbf{u}[k] + \Delta \mathbf{u}[k]
    \end{aligned}
    \right .
\end{equation}

If the system is sufficiently well behaved, the LPV counterpart provides a good approximation to the original system dynamics, hence why employing a quasi-LPV model can be a potential representative for nonlinear systems alike, considering $\Theta [k] = (\mathbf{x}[k]^T,\mathbf{u}[k]^T)^T$.
This means that the full scheduling function of the quasi-LPV system is the Jacobian.
In a black-box scenario, the identification problem consists of identifying the nonlinear system Jacobian and employing it to predict the system.

\mbox{}

\noindent\textit{\textbf{Example}:} Representing a system in LPV form is a matter of context.
For instance, take a simple mass-spring system governed by the following differential equations:
\begin{equation}
\begin{aligned}
    \dot{x} &= v_x\\
    \dot{v}_x &= \frac{F - kx - bv_x}{m}
\end{aligned}
\end{equation}
where $x$ is the load position, $v_x$ is the load velocity, $m$ is the load mass, $k$ is the spring constant, $F$ is any exogenous force exerted in the system, and $b$ is the viscous friction constant.

The model is linear by itself, but if some considerations about the problem are made, it can be regarded as an LPV system.
For instance, suppose the viscous friction constant varies over time and is assumed to be not known with exactitude.
Then, if one considers $\Theta = b$, and $m = k = 1$ for this system:
\begin{equation}
\begin{aligned}
    \mathbf{A}(\Theta) &= 
    \begin{bmatrix}
      0 & 1\\
      -1 & -\Theta
    \end{bmatrix}\\
    \mathbf{B}(\Theta) &=
    \begin{bmatrix}
        0\\
        1
    \end{bmatrix}
\end{aligned}
\end{equation}

Another convenient example of a (quasi)-LPV system is the Van der Pol oscillator:
\begin{equation}
\begin{aligned} \label{eq:sys:VandePol}
  \dot{x}_1 &= x_2\\
  \dot{x}_2 &= \mu(1-x_1^2)x_2 - x_1 + u
\end{aligned}
\end{equation}

The Van der Pol oscillator is a fully nonlinear system; however, if $\mu = 1$ and $\Theta = x_1^2$, then it can be represented as an LPV system with:
\begin{equation}
\begin{aligned}
    \mathbf{A}(\Theta) &= 
    \begin{bmatrix}
      0 & 1\\
      -1 & 1 -\Theta
    \end{bmatrix}\\
    \mathbf{B}(\Theta) &=
    \begin{bmatrix}
        0\\
        1
    \end{bmatrix}
\end{aligned}
\end{equation}

Since a function of the state was placed as a parameter, the system in question is referred to as  quasi-LPV. However, many of the analysis tools remain the same for both LPV and quasi-LPV.

\subsection{LPV System Identification} \label{sec:lpv_ident}

Consider an arbitrary LPV system in the form \eqref{eqn:lpvs}.
Also, assume that both matrix functions $\mathbf{A}(\Theta)$ and $\mathbf{B}(\Theta)$ have the following form for the purpose of black-box system identification:
\begin{align}\label{eqn:scheduling_function}
\left \{ 
\begin{aligned}
    \mathbf{A}(\boldsymbol{\Theta}[k]) &= \mathbf{A}_0 + \sum_{i=1}^{N_f}\mathbf{A}_i\phi_i(\Theta[k])\\
    \mathbf{B}(\boldsymbol{\Theta}[k]) &= \mathbf{B}_0 + \sum_{i=1}^{N_f}\mathbf{B}_i\psi_i(\Theta[k])
    \end{aligned} \right .
\end{align}
where $\phi_i$ and $\psi_i$ (with $i \in \{1,2,\hdots,N_f\}$ are scalar functions, the so-called basis functions,  with $N_f$ being the number of different basis functions to compute the scheduling function of the LPV system, through linear combination. This means that the scheduling functions $\mathbf{A}(\boldsymbol{\Theta})$ and $\mathbf{B}(\boldsymbol{\Theta})$ are linear in the parameters, whereas the weights $\mathbf{A}_i$ and $\mathbf{B}_i$ are constants for $i \in \{0,1,\hdots,N_f\}$.

Considering the scheduling function candidates in Eqn. \eqref{eqn:scheduling_function}, the LPV system \eqref{eqn:lpvs} becomes:
\begin{equation} \label{eqn:full_lpv}
        \mathbf{x}[k+1] = \left(\mathbf{A}_0 + \sum_{i=1}^{N_f}\mathbf{A}_i\phi_i(\Theta[k])\right)\mathbf{x}[k] + \left(\mathbf{B}_0 + \sum_{i=1}^{N_f}\mathbf{B}_i\psi_i(\Theta[k])\right)\mathbf{u}[k] 
\end{equation}

An alternative way to present the LPV system above is as follows:
\begin{equation} \label{eq:LPV:Kronecker}
 \mathbf{x}[k+1] = \mathbf{W_A}\Bigl (\underbrace{\boldsymbol{\phi}(\Theta[k]) \otimes \mathbf{x}[k]}_{\text{features}}\Bigr)    
   + \mathbf{W_B}\Bigl (\underbrace{\boldsymbol{\psi}(\Theta[k]) \otimes \mathbf{u}[k]}_{\text{features}}\Bigr )
 \end{equation}
 where:
\begin{equation} \label{eq:LPV:Kronecker:matrices}
\left \{
\begin{aligned}
 \mathbf{W_A} &=
 \begin{pmatrix}
     \mathbf{A}_0 & \mathbf{A}_1 & \cdots & \mathbf{A}_{N_f}
 \end{pmatrix}\\
 \mathbf{W_B} &= \begin{pmatrix}
     \mathbf{B}_0 & \mathbf{B}_1 & \cdots & \mathbf{B}_{N_f}
 \end{pmatrix} \\
 \boldsymbol{\phi} &=
 \begin{pmatrix}
     1 & \phi_1(\Theta[k]) & \phi_2(\Theta[k]) & \cdots & \phi_{N_f}(\Theta[k])
 \end{pmatrix}^T \\
 \boldsymbol{\psi} &=
 \begin{pmatrix}
     1 & \psi_1(\Theta[k]) & \psi_2(\Theta[k]) & \cdots & \psi_{N_f}(\Theta[k])
 \end{pmatrix}^T
    \end{aligned}
        \right .
\end{equation}
in which $\otimes$ is the Kronecker product operator.
The Kronecker product of two column vectors represents the second operand being replicated $N_f$ times, with each instance being multiplied by the corresponding $\phi_i$ or $\psi_i$.
The column of the resulting \textit{feature} vector has dimension $N_f \times n_s$ for the states and $N_f \times n_{in}$ for the inputs, respectively, with $n_s$ being the number of states and $n_{in}$ being the number of inputs. 

Regarding the Van der Pol system \eqref{eq:sys:VandePol} with $\mu=1$, it can be written in the following alternative form using the Kronecker product, referring to a continuous-time counterpart of  the definition in \eqref{eq:LPV:Kronecker}:
\begin{equation}
   \left \{ \begin{aligned}
        \boldsymbol{\phi}(\Theta) &= 
        \begin{bmatrix}
         1 \\ \Theta   
        \end{bmatrix}\\
        \mathbf{A_0} &=
        \begin{bmatrix}
            0 & 1\\
            -1 & 1
        \end{bmatrix}, &\quad &
        \mathbf{A_1} &=
        \begin{bmatrix}
            0 & 0\\
            0 & -1
        \end{bmatrix}\\
        \mathbf{B_0} &=
        \begin{bmatrix}
        0\\
        1
    \end{bmatrix}
    \end{aligned} \right .
\end{equation}

Since we do not know the LPV model parameters a priori, the problem of identifying an LPV reduces to identifying the parameters $\mathbf{W_A}$ and $\mathbf{W_B}$ of the scheduling function.
There are two main strategies for identifying an LPV system:
\begin{itemize}
    \item \textbf{Global Identification}: The system is identified holistically from the whole dataset by performing identification considering the parameters, current states, and control inputs as features and the following states as output.
    This strategy is the same as identifying any model linear on the parameters, performing least squares on weights that multiply a set of features, leading to the curse of dimensionality, which results from an exponential number of features. Mathematically, the features result from the Kronecker products in Eq. \eqref{eq:LPV:Kronecker}, which is the product between the states (or inputs) and the parameters. In contrast, the weights are the matrices $\mathbf{W_A}$  and $\mathbf{W_B}$.
    If global LPV were performed in a system with a large number of states (\textit{e.g.}, any PDE-derived system), a combinatorial explosion would naturally ensue.
    
    \item \textbf{Local Identification}: Provided the parameters can be made constant for an arbitrarily long period of time, one can identify an LPV system through a collection of LTI systems.
    Whenever a parameter is constant, the LPV system is an LTI system.
    Therefore, the local identification can be performed in two steps:
    \begin{enumerate}
        \item Obtain a family of LTI systems, each associated with a constant parameter instance $\Theta_i$ for the $i^{\text{th}}$ LTI system identified.
        \item Identify the scheduling function \eqref{eqn:scheduling_function} with the LTI parameters as output and the corresponding $\Theta_i$ as input, with a dataset the size of the number of LTI systems identified.
    \end{enumerate}
    One can only perform local identification in slow systems, where a constant parameter can be assumed.
\end{itemize}

\subsection{Proper Orthogonal Decomposition for LPV Systems}  \label{sec:pod_lpv}

Proper Orthogonal Decomposition (POD) is a model order reduction method \cite{Jordanou2023} performed in the context of Dynamic Mode Decomposition (DMD) \cite{DMDc}, as Section \ref{sec:dmdc} shows.
POD consists of obtaining a linear transformation that optimally maps the full states of a system into a reduced state \cite{Jordanou2023}.
The mapping naturally occurs through the Singular Value Decomposition (SVD) of a matrix $\mathbf{X}$ containing snapshots of the state at given time instants:
\begin{equation}
    \mathbf{W}\mathbf{\Sigma}\mathbf{V}^{T} = \mathbf{X}
\end{equation}
where each column of $\mathbf{X}$ is the system state at a given time.

The singular value decomposition orders the left singular values by how much the state in a resulting transformation contributes to the system output.
Therefore, the transformation $\mathbf{T}$ is only the first few columns of the left-singular vectors matrix $\mathbf{W}=[\mathbf{T}~\widetilde{\mathbf{T}}]$, according to the dimension of the reduced state.
Thus, obtaining the mapping:
\begin{equation}
    \mathbf{x}[k] = \mathbf{T}\mathbf{z}[k]
\end{equation}
where $\mathbf{x}$ are the states in the full dimension and $\mathbf{z}$ are the states in the reduced order space.

Since the singular vectors matrix has the property $\mathbf{T}^T\mathbf{T} = \mathbf{I}$, obtaining a reduced linear system from $\mathbf{x}[k+1] = \mathbf{A}\mathbf{x}[k] + \mathbf{B}\mathbf{u}[k]$ is trivial:
\begin{equation}
   \mathbf{z}[k+1] = \mathbf{T}^T\mathbf{AT}\mathbf{z}[k] + \mathbf{T}^T\mathbf{B}\mathbf{u}[k]
\end{equation}
Notice that matrices such as $\mathbf{T}^T\mathbf{AT}$ can be calculated offline or, in other words, before the simulation run.

The matter with nonlinear systems is not as straightforward:
\begin{equation}
    \mathbf{z}[k+1] = \mathbf{T}^T\mathbf{f}(\mathbf{Tz}[k],\mathbf{u[k]})
\end{equation}
since, besides the reduction being performed, the computation remains of the scale of the full order model because of the $\mathbf{T}^T\mathbf{f}(\cdot)$ term.
Countermeasures to reduce computation, such as Discrete Empirical Interpolation (DEIM) \cite{Chatu2010}, are not discussed in this work.

For LPV systems such as \eqref{eq:LPV:Kronecker}, the reduction has the following form:
\begin{equation}
 \mathbf{z}[k+1] = \mathbf{T}^T\mathbf{W_A}\Bigl(\boldsymbol{\phi}(\Theta[k]) \otimes \mathbf{T}\mathbf{z}[k]\Bigr) 
     + \mathbf{T}^T\mathbf{W_B}\Bigl(\boldsymbol{\psi}(\Theta[k]) \otimes \mathbf{u}[k]\Bigr)   
\end{equation}

The Kronecker product obeys the following property given four arbitrary matrices with matching dimensions:
\begin{equation}
    \mathbf{AB} \otimes \mathbf{CD} = (\mathbf{A} \otimes \mathbf{C})(\mathbf{B} \otimes \mathbf{D}),
\end{equation}
which is known as the \textit{Kronecker mixed-product property} \cite{Zhang2013}.

By using this property, we can rewrite the LPV system \eqref{eq:LPV:Kronecker} as follows:
\begin{equation}
\begin{aligned}
 \mathbf{z}[k+1] &= \mathbf{T}^T\mathbf{W_A}(\mathbf{I} \otimes \mathbf{T})\bigl(\boldsymbol{\phi}(\Theta[k]) \otimes \mathbf{z}[k]\bigr)    + \mathbf{T}^T\mathbf{W_B}\bigl (\boldsymbol{\psi}(\Theta[k]) \otimes \mathbf{u}[k]\bigr)   \\
    &= \widetilde{\mathbf{W}}_{\mathbf{A}}\bigl(\boldsymbol{\phi}(\Theta[k]) \otimes \mathbf{z}[k]\bigr)   +\widetilde{\mathbf{W}}_{\mathbf{B}}\bigl(\boldsymbol{\psi}(\Theta[k]) \otimes \mathbf{u}[k]\bigr) 
 \end{aligned} \label{z:reduced:dynamics}
\end{equation}
This way, the matrix  $\widetilde{\mathbf{W}}_{\mathbf{A}}=\mathbf{T}^T\mathbf{W_A}(\mathbf{I} \otimes \mathbf{T})$ can be calculated offline, which means that LPV and quasi-LPV systems benefit fully from the POD method. Notice that the reduced-order system dynamics in \eqref{z:reduced:dynamics} is analogous to the full-state system dynamics in \eqref{eq:LPV:Kronecker}.

\section{Dynamic Mode Decomposition with Control (DMDc)}\label{sec:dmdc}

Consider the following linear state-space model with state $\mathbf{x}$ and control input $\mathbf{u}$:
\begin{equation}
    \mathbf{x}[k+1] = \mathbf{A}\mathbf{x}[k] + \mathbf{B}\mathbf{u}[k]
        \label{eq:DMCc:full-state:dyn}
\end{equation}
where the number of states $n_s$ of $\mathbf{x}[k]$ is large enough so that the matrix $\mathbf{A}$ is difficult to compute directly, and $n_u$ is the number of control inputs.

Instead of directly solving the least squares problem, the Dynamic Mode Decomposition addresses the rank-limited Procrustes problem \cite{DMDc}:
\begin{subequations} \label{eqn:procustes}
\begin{align}
    \min_{\mathbf{A},\mathbf{B}} ~&\| \mathbf{Y} - (\mathbf{A}\mathbf{X} + \mathbf{B}\mathbf{U})\|\\
    \text{s.t.} ~~&  \text{rank}(\left[\mathbf{A},\mathbf{B}\right]) = r
\end{align}
\end{subequations}
where:
\begin{equation}
\left \{ \begin{aligned}
    \mathbf{X} &=
    \begin{pmatrix}
    \mathbf{x}[0] & \mathbf{x}[1] & \cdots & \mathbf{x}[N-1] 
    \end{pmatrix}\\
    \mathbf{U} &=
    \begin{pmatrix}
    \mathbf{u}[0] & \mathbf{u}[1] & \cdots & \mathbf{u}[N-1] 
    \end{pmatrix}\\
    \mathbf{Y} &=
    \begin{pmatrix}
    \mathbf{x}[1] & \mathbf{x}[2] & \cdots & \mathbf{x}[N] 
    \end{pmatrix}
\end{aligned} \right . \label{eq:def:X-U-Y}
\end{equation}

DMDc solves the least squares problem, limiting the solution's rank to a given value $r$.
One obtains this solution by performing SVD on the concatenation of $\mathbf{X}$ and $\mathbf{U}$:
\begin{equation}
    \mathbf{W}\mathbf{\Sigma}\mathbf{V}^T =
    \begin{bmatrix}
        \mathbf{X}\\
        \mathbf{U}
    \end{bmatrix}
\end{equation}
where $\mathbf{W}$, the left singular vectors matrix, has dimension $n \times N$, $\mathbf{\Sigma}$, the singular values matrix, has dimension $N \times N$, $\mathbf{V}$ has dimension $N \times N$; further, $n=n_s+n_u$ is the number of states and control inputs and $N$ is the number of training examples/snapshots contained inside the matrix.

The solution to the standard least squares problem is expressed as \cite{DMDc}:
\begin{equation}
    \begin{bmatrix}
        \mathbf{A} & \mathbf{B} 
    \end{bmatrix} = \mathbf{Y}\mathbf{V}\mathbf{\Sigma}^{-1}\mathbf{W}^T
\end{equation}

The singular value matrix $\mathbf{\Sigma}$ is a good indicator of whether or not the least squares problem is well posed.
As a diagonal matrix, the inversion is merely the reciprocal of each element.
If a singular value is too small, the inversion $\mathbf{\Sigma}^{-1}$ can result in values close to infinity.
Regularization is a popular solution to bad training data conditioning and, in the SVD framework, it can be included in the problem as follows:
\begin{equation}
    \begin{bmatrix}
        \mathbf{A} & \mathbf{B} 
    \end{bmatrix} = \mathbf{Y}\mathbf{V}\mathbf{\Sigma^{reg}}\mathbf{W}^T
\end{equation}
where $\mathbf{A}$ corresponds to the first $n_s$ columns of the result of the right-hand side operation, and $\mathbf{B}$ is composed of the remaining $n_u$ columns.
Each diagonal element of $\mathbf{\Sigma^{\mathbf{reg}}}$ is defined as follows:
\begin{equation}
   \sigma^{\mathbf{reg}}_i = \frac{\sigma_i}{\sigma_i^2 + \lambda^2}
\end{equation}
where $\lambda$ is the regularization parameter and $\sigma_i$ is the $i^{th}$ largest singular value.
This equation shows how the regularization term avoids undesirably large values by having the regularization parameter dominate over the small singular value.

To constrain the rank $r$ of the solution, we consider only the first $r$ columns of $\mathbf{W}$ and $\mathbf{V}$, and keep only the largest $r$ singular values of $\mathbf{\Sigma}$, obtaining $\mathbf{W}_r$, $\mathbf{V}_r$, and $\mathbf{\Sigma}_r^{\mathbf{reg}}$.

With DMDc, we perform model-order reduction without computing the full matrices $\mathbf{A}$ and $\mathbf{B}$.
We define a new SVD from $\mathbf{Y}$, performing the same rank-$r$ truncation as the previously described SVD:
\begin{equation} \label{eqn:pod_in_dmd}
\mathbf{W}_{\mathbf{y}r}\mathbf{\Sigma}_{\mathbf{y}r}\mathbf{V}_{\mathbf{y}r}^T \approx \mathbf{Y}
\end{equation}

Then, we perform POD for the definitions of $\mathbf{A}$ and $\mathbf{B}$ according to DMDc \cite{DMDc} with transformation matrix $\mathbf{W}_{\mathbf{y}r}$:
\begin{equation} \label{eqn:dmd2}
\left \{
\begin{aligned}
    \widetilde{\mathbf{A}} &= \mathbf{W}_{\mathbf{y}r}^T\overbrace{\mathbf{Y}\mathbf{V}_{r}\mathbf{\Sigma}_{r}^{\mathbf{reg}}
                \mathbf{W}_{r,1}^T}^{\mathbf{A}}\mathbf{W}_{\mathbf{y}r}\\
    \widetilde{\mathbf{B}} &= \mathbf{W}_{\mathbf{y}r}^T\underbrace{\mathbf{Y}\mathbf{V}_{r}
         \mathbf{\Sigma}_{r}^{\mathbf{reg}}\mathbf{W}_{r,2}^T}_{\mathbf{B}}
\end{aligned}
         \right .
\end{equation}
obtaining the reduced-order matrices $\widetilde{\mathbf{A}}$ and $\widetilde{\mathbf{B}}$, with $\mathbf{W}_{r,1}$ corresponding to the rows of $\mathbf{W}_{r}$ related to $\mathbf{A}$, and $\mathbf{W}_{r,2}$ corresponding to the rows of $\mathbf{W}_{r}$ related to $\mathbf{B}$, namely
\begin{equation*}
   \mathbf{W}_r = 
       \begin{bmatrix}
            \mathbf{W}_{r,1} \\
            \mathbf{W}_{r,2}
        \end{bmatrix}
\end{equation*}

The reduced-order model can now be expressed as:
\begin{equation}
\mathbf{z}[k+1] = \widetilde{\mathbf{A}}\mathbf{z}[k] 
         + \widetilde{\mathbf{B}}\mathbf{u}[k] 
\end{equation}

The non-zero eigenvalues of $\widetilde{\mathbf{A}}$ are also eigenvalues of $\mathbf{A}$ \cite{Htu2014}, and an eigenvector $\boldsymbol{\phi}$ of $\mathbf{A}$ can be obtained given an eigenvalue-eigenvector pair $(\lambda,\boldsymbol{\omega})$ of $\widetilde{\mathbf{A}}$ as follows:
\begin{equation}
    \boldsymbol{\phi} = \frac{1}{\lambda}\mathbf{Y}\mathbf{V}\mathbf{\Sigma}_r^{\mathbf{reg}}\boldsymbol{\omega}
\end{equation}

When applying DMDc to noisy data \cite{Dawson2016}, it is important to recognize that DMD can effectively manage output data contaminated by white noise, as the method relies on Least Squares. However, noise present in the input data can introduce bias errors during system identification.
Some studies, such as \cite{Dawson2016}, propose strategies for adapting DMD to filter out noise in both the input and state data. However, our research does not address noise in the experiments, and we have considered the treatment of noise for future studies.

In summary, DMDc is a faster way to compute eigenvalues and eigenvectors of a large-scale system without having to compute the matrix itself \cite{Htu2014}.
The algorithm was designed for situations where $n \gg N$ ($r = N$), where $n$ is so large that gathering $n$ samples is impeditive (\textit{e.g.}, grid-based PDE simulations, as an oil reservoir).

\section{Identification of Large-Scale LPV Systems} \label{sec:dmd_lpv}

The idea developed in this work is based on \cite{Gosea2021}, which shows how to apply DMD to systems that are nonlinear but linear in the parameters.
The method is the same as DMDc, with the nonlinear features added to the Procrustes problem's input \eqref{eqn:procustes_lpv}.
We reformulate the same idea in the context of LPV system identification, both locally and globally.

\subsection{Global LPV System Identification} \label{subsec:Glb-LPV}

The least squares cost function for LPV cases is:
\begin{equation} \label{eqn:procustes_lpv}
\begin{aligned}
    \min_{\mathbf{W_A},\mathbf{W_B}} ~& \| \mathbf{Y} - (\mathbf{W_A}\mathbf{X_P} + \mathbf{W_B}\mathbf{U_P})\|\\
    \text{s.t.} ~~& \text{rank}(\left[\mathbf{A},\mathbf{B}\right]) = r
\end{aligned}
\end{equation}
where $\mathbf{X_P}$ and $\mathbf{U_P}$ are matrices that have their columns defined by the Kronecker product $\otimes$ between each corresponding column of the parameter data matrix $\mathbf{P_x}$ and state data matrix $\mathbf{X}$ for $\mathbf{X_P}$; and $\mathbf{P_u}$ with $\mathbf{U}$ for $\mathbf{U_P}$.
The parameter data matrix $\begin{bmatrix}\mathbf{P_x} & \mathbf{P_u}\end{bmatrix}$ contains the features resulting from the nonlinear mapping of the parameters, with each column corresponding to a given instant in time (note that $\boldsymbol{\phi}[k] \coloneqq \boldsymbol{\phi}(\Theta [k])$), with the analog being valid for $\psi$:
\begin{equation} \label{eq:def:Px-Pu}
 \left \{ 
\begin{aligned}
  \mathbf{P_x} &=
    \begin{pmatrix}
        \boldsymbol{\phi}[0] & \boldsymbol{\phi}[1] & \cdots & \boldsymbol{\phi}[N-1]
    \end{pmatrix}\\
     \mathbf{P_u} &=
    \begin{pmatrix}
        \boldsymbol{\psi}[0] & \boldsymbol{\psi}[1] & \cdots & \boldsymbol{\psi}[N-1]
    \end{pmatrix}
\end{aligned} \right .
\end{equation}
and, therefore, $\mathbf{X_p}$ and $\mathbf{U_p}$ have the following form:
\begin{equation} \label{eq:def:Xp-Up}
 \left \{ 
\begin{aligned}
    \mathbf{X_P} &=
    \begin{pmatrix}
        \boldsymbol{\phi}[0] \otimes \mathbf{x}[0] & \boldsymbol{\phi}[1] \otimes \mathbf{x}[1] & \cdots & \boldsymbol{\phi}[N-1] \otimes \mathbf{x}[N-1]
    \end{pmatrix}\\
     \mathbf{U_P} &=
    \begin{pmatrix}
        \boldsymbol{\psi}[0] \otimes \mathbf{u}[0] & \boldsymbol{\psi}[1] \otimes \mathbf{u}[1] & \cdots & \boldsymbol{\psi}[N-1]\otimes \mathbf{u}[N-1]
    \end{pmatrix}
\end{aligned} \right .
\end{equation}

For the global method, the application of DMD is straightforward, as the least squares solution for such LPV identification problems is:
\begin{equation}
    \begin{bmatrix}
        \mathbf{W_A} & \mathbf{W_B} 
    \end{bmatrix} = \mathbf{Y}\mathbf{V}\mathbf{\Sigma^{reg}}\mathbf{W}^T
\end{equation}
with $\mathbf{Y}$ defined as in \eqref{eq:def:X-U-Y}, and 
considering the following SVD operation and the definitions of Section \ref{sec:dmdc}:
\begin{equation}
    \mathbf{W}\mathbf{\Sigma}\mathbf{V}^T =
    \begin{bmatrix}
        \mathbf{X_P}\\
        \mathbf{U_P}
    \end{bmatrix}
\end{equation}

The reduction follows \eqref{eqn:pod_in_dmd} exactly as in DMDc.
However, the next step must be performed differently.
By performing Model Order Reduction with $\eqref{eqn:pod_in_dmd}$ in an LPV system of the same type as \eqref{eqn:full_lpv}, one obtains the following reduced system:
\begin{multline} \label{eqn:mor_lpv}
        \mathbf{z}[k+1] = \left(\mathbf{W}_{\mathbf{y}r}^T\mathbf{A}_0\mathbf{W}_{\mathbf{y}r} + \sum_{i=1}^{N_f}\mathbf{W}_{\mathbf{y}r}^T\mathbf{A}_i\mathbf{W}_{\mathbf{y}r}\phi_i(\Theta[k])\right)\mathbf{z}[k] \\ 
          + \mathbf{W}_{\mathbf{y}r}^T\left(\mathbf{B}_0 + \sum_{i=1}^{N_f}\mathbf{B}_i\psi_i(\Theta[k])\right)\mathbf{u}[k] 
\end{multline}

One can see that the reduced matrix for $\mathbf{W_{B}}$ is $\widetilde{\mathbf{W}}_{\mathbf{B}} = \mathbf{W}_{\mathbf{y}r}^T\mathbf{W_B}$, making the definition of the control weights similar to the DMDc case for linear systems. On the other hand, the reduced matrix $\widetilde{\mathbf{W}}_{\mathbf{A}}$ for $\mathbf{W_A}$  is obtained differently.

The compact form of the LPV places the reduced system in the following form:
\begin{multline}
\mathbf{z}[k+1] = \mathbf{W}_{\mathbf{y}r}^T\mathbf{W_A}\Bigl(\boldsymbol{\phi}(\Theta[k]) \otimes \mathbf{W}_{\mathbf{y}r}\mathbf{z}[k]\Bigr)   \\  
 + \mathbf{W}_{\mathbf{y}r}^T\mathbf{W_B}\Bigl(\boldsymbol{\psi}(\Theta[k]) \otimes \mathbf{u}[k]\Bigr)    
\end{multline}

By considering the mixed-product property of the Kronecker product, we obtain the equivalent representation:
\begin{multline}
\mathbf{z}[k+1] = \mathbf{W}_{\mathbf{y}r}^T\mathbf{W_A}\Bigl(\mathbf{I}_p \otimes \mathbf{W}_{\mathbf{y}r}\Bigr)\Bigl(\boldsymbol{\phi}(\Theta[k]) \otimes \mathbf{z}[k]\Bigr) 
  \\ + \mathbf{W}_{\mathbf{y}r}^T\mathbf{W_B}\Bigl(\boldsymbol{\psi}(\Theta[k]) \otimes \mathbf{u}[k]\Bigr)    
\end{multline}
where $\mathbf{I}_p$ is an identity matrix of the same dimension as the number of  parameter functions.

The newly obtained Kronecker product $(\mathbf{I}_p \otimes \mathbf{W}_{\mathbf{y}r})$ represents directly the post multiplication of each block element $\mathbf{A}_i$ by $\mathbf{W}_{\mathbf{y}r}$.

The result of this discussion is the synthesis of the reduced form of the weights of the DMD over the LPV system: 
\begin{equation}
\left \{
\begin{aligned} \label{eqn:dmdc_lpv}
    \widetilde{\mathbf{W}}_{\mathbf{A}} &= \mathbf{W}_{\mathbf{y}r}^T\mathbf{Y}\mathbf{V}_{r}\mathbf{\Sigma}_r^{\mathbf{reg}}\mathbf{W}_{r,1}^T\left(\mathbf{I}_p \otimes \mathbf{W}_{\mathbf{y}r}\right)\\
    \widetilde{\mathbf{W}}_{\mathbf{B}} &= \mathbf{W}_{\mathbf{y}r}^T\mathbf{Y}\mathbf{V}_r\mathbf{\Sigma}_{r}^{\mathbf{reg}}\mathbf{W}_{r,2}^T
\end{aligned}
  \right .
\end{equation}

The methods for obtaining the dynamic modes and approximate original eigenvectors are precisely as in DMDc, with the detail that one can calculate the eigenvalues and eigenvectors related to the contribution of a specific block $\mathbf{A}_i$ related to feature map $\phi_i(\Theta [k])$ of $\mathbf{W_A}$.

\subsection{Local LPV Identification}

Local system identification entails applying the system identification procedure multiple times.
 To perform local identification in the DMD-LPV context, we first obtain the POD transformation matrix $\mathbf{W}_{\mathbf{y}r}$ with the entire dataset of state snapshots.
Then, the reduced weights are identified directly by performing POD on the full matrices $\mathbf{A}(i)$ and $\mathbf{B}(i)$, which are obtained beforehand.
The equation that defines the identification of $\widetilde{\mathbf{W}}_{\mathbf{A}}$ and $\widetilde{\mathbf{W}}_{\mathbf{B}}$ in this case is:
   \begin{equation}
   \left \{
\begin{aligned} \label{eqn:local_dmdlpv}
     \mathbf{W}_{\mathbf{y}r}^T\mathbf{A}(i)\mathbf{W}_{\mathbf{y}r} &=\widetilde{\mathbf{W}}_{\mathbf{A}}(\phi(\boldsymbol{\Theta}_i) \otimes \mathbf{I_s})\\
     \mathbf{W}_{\mathbf{y}r}^T\mathbf{B}(i) &=\widetilde{\mathbf{W}}_{\mathbf{B}}(\psi(\boldsymbol{\Theta}_i) \otimes \mathbf{I_{in}})
\end{aligned}
 \right .
\end{equation}
where $\mathbf{I_s}$ is the identity matrix on the same dimension as the reduced state, and $\mathbf{I_{in}}$ is the identity matrix on the same dimension as the number of system inputs.
Although the system outputs are matrices, least squares can be performed normally for both reduced weights.
Recovering the original system involves the same steps as the global LPV identification and DMDc cases.

Note that there are two possible approaches for this identification:
\begin{itemize}
    \item Identify $\mathbf{W}_{\mathbf{y}r}$ and the LTI systems separately, and then reduce the matrices dimension during the identification of the LPV parameters.
    \item Use $\mathbf{W}_{\mathbf{y}r}$ to identify each LTI system by mapping the states to the reduced dimension, so no identification is performed in the full space.
\end{itemize}

In both cases, the reduced weights can be identified by solving the Procrustes problem, which makes this method more efficient than global identi\-fi\-cation---the global case uses a Kronecker product between the parameters and the full state inputs as input matrix to the SVD.
The proposal to identify each LTI system in the latent space obeys the DMD logic of noninvasive MOR identification since every model training is not performed in the full state space.
In fact, every linear system is identified in the context of the full LPV model latent space, and the POD transformation $\mathbf{W}_{\mathbf{y}r}$ serves to map the reduced order states into the full order states in the context of the LPV model.

\subsection{Local and Global Identification Procedures}

To summarize the discussion in this section, the global identification of a DMD-LPV model for a given system is formalized in Algorithm \ref{alg:global:LPV}.

\begin{algorithm}
  \caption{Global DMD-LPV Identification\label{alg:global:LPV}}
  \begin{algorithmic}[1] 
    \Inputs{State data $\mathbf{X}$, Input data $\mathbf{U}$, Parameter data $\mathbf{P}$, Output data $\mathbf{Y}$ (state forwarded 1t time step), Procrustes rank $r_{pr}$, POD rank $r_{pod}$, Regularization parameter ($\lambda$)}
    \Initialize{Obtain $\mathbf{P_x}$ and $\mathbf{P_u}$ from $\mathbf{P}$ according to Eqn. \eqref{eq:def:Px-Pu}}, 
    \For{$k = 1$ to $N$}
    \State $\mathbf{X_p}[:,k] \gets \mathbf{P_x}[:,k] \otimes \mathbf{X}[:,k]$
    \State $\mathbf{U_p}[:,k] \gets \mathbf{P_u}[:,k] \otimes \mathbf{U}[:,k]$   
    \EndFor
    \State $\mathbf{INPUT_{svd}} \gets $ Vertical concatenation of $\mathbf{X_p}$ and $\mathbf{U_p}$.
    
    \State $\mathbf{W}_r,\mathbf{s}_r,\mathbf{V}_r \gets svd(\mathbf{INPUT_{svd}})$. SVD is truncated to rank $r = r_{pr}$.
    
    \State $\mathbf{\Sigma}_{r}^{\mathbf{reg}} \gets$ perform the operation: $\mathbf{s}_r \oslash( \mathbf{s}_r\odot \mathbf{s}_r + \lambda^2\mathbf{1})$. ($\odot$: element-wise product, $\oslash$: element-wise division.)
    
    \State $\mathbf{W}_{\mathbf{y}r} \gets$ left-singular vectors of $svd(\mathbf{Y})$ (truncated to rank $r_{pod}$).
    \State $\mathbf{W}_{r,1} \gets $ First rows of $\mathbf{W_r}$ corresponding to $n_s \times n_f$. (Number of states times number of parameter features in state computation).
    \State $\mathbf{W}_{r,2} \gets $ Rows of $\mathbf{W}_r$ that are not in $\mathbf{W}_{r,1}$
    \State Compute weights with Eqn. \eqref{eqn:dmdc_lpv}.
  \end{algorithmic}
\end{algorithm}

The solution of the least squares problem would require finding the pseudo-inverse of a matrix with $n_f \times n_s$ features to compute the weights of the LPV model for $n_s$ outputs, with a global identification method.
The assumption for a large-scale LPV system is that $n_s$ is a huge number, and multiplying it by $n_f$ might make the LPV problem very difficult to solve.
This is pertinent for systems with an $n_s$ dimension as a function of $n_f$.
A quick assessment of the algorithm reveals that
the rank truncation reduces the problem to compute a $r_{pr}-$rank approximation for the pseudo-inverse, which is considerably cheaper than the alternative.
Also, the POD reduction leads to the calculation of fewer weights.

In turn, the local identification of a DMD-LPV system follows Algorithm 2.

\begin{algorithm}
  \caption{Local DMD-LPV Identification \label{alg:local:LPV}}
  \begin{algorithmic}[1] 
    \Inputs{An arbitrary number of values for $\mathbf{p}$. For each value $\mathbf{p}_i$, a corresponding state matrix $\mathbf{X}_i$, input matrix $\mathbf{U}_i$ and output-state matrix $\mathbf{Y}_i$, with each matrix obtained by exciting the LPV frozen in value $\mathbf{p}_i$ with $\mathbf{U}_i$. Rank $r$.}
    \Initialize{$n_{vp} \gets $ number of values configuration used for $\mathbf{p}$.\\ $\mathbf{Y_{tot}} \gets$ Concatenation of all output matrices $\mathbf{Y}_i$.}
    \State $\mathbf{W}_{\mathbf{y}r} \gets$ left-singular vector of $svd(\mathbf{Y_{tot}})$.
    \For{$i = 1$ to $n_{vp}$}
    \State Identify discrete LTI system $\mathbf{x}[k+1] = \mathbf{A}(i)\mathbf{x}[k] + \mathbf{B}(i)\mathbf{u}[k]$ with fixed $\mathbf{p}$ and corresponding $\mathbf{X}_i,\mathbf{U}_i$ and $\mathbf{Y}_i$ (Least Squares or Procrustes problem).
    \State Store matrices $\mathbf{A}(i)$ and $\mathbf{B}(i)$ of LTI system.
    \EndFor
    \State $\mathbf{\widetilde{W}_A},\mathbf{\widetilde{W}_B} \gets$ Find reduced weights by solving the $r$-rank Procrustes problem for the system $\eqref{eqn:local_dmdlpv}$, with every parameter as input and every corresponding LTI weight as output (reduced by the POD transformation).
  \end{algorithmic}
\end{algorithm}

Algorithm \ref{alg:local:LPV} has the issue of identifying the LTI systems on the original dimension, which defeats the purpose of the DMD non-intrusive MOR.
As an alternative, we propose using the POD identification from $\mathbf{Y_{tot}}$ to obtain a latent space and identify the LTI systems in it, which is formalized in Algorithm \ref{alg:local:LPV:B}.

\begin{algorithm}
  \caption{Local Non-Intrusive DMD-LPV Identification \label{alg:local:LPV:B}}
  \begin{algorithmic}[1] 
    \Inputs{An arbitrary number of values for $\mathbf{p}$. For each value $\mathbf{p}_i$, a corresponding state matrix $\mathbf{X}_i$, input matrix $\mathbf{U}_i$ and output-state matrix $\mathbf{Y}_i$, with each matrix obtained by exciting the LPV frozen in value $\mathbf{p}_i$ with $\mathbf{U}_i$. Rank $r$.}
    \Initialize{$n_{vp} \gets $ number of values configuration used for $\mathbf{p}$.\\ $\mathbf{Y_{tot}} \gets$ Concatenation of all output matrices $\mathbf{Y}_i$.}
    \State $\mathbf{W}_{\mathbf{y}r} \gets$ left-singular vectors of $svd(\mathbf{Y_{tot}})$ ($r$-rank truncation).
    \For{$i = 1$ to $n_{vp}$}
    \State Identify discrete LTI system $\mathbf{x}[k+1]=\mathbf{\widetilde{A}}(i)\mathbf{x}[k] + \mathbf{\widetilde{B}}(i)\mathbf{u}[k]$ with fixed $\mathbf{p}$ and corresponding $\mathbf{W}_{\mathbf{y}r}^T\mathbf{X}_i,\mathbf{U}_i$ and $\mathbf{W}_{\mathbf{y}r}^T\mathbf{Y}_i$ (Least Squares or Procrustes problem).
    \State Get matrices $\mathbf{\widetilde{A}}(i)$ and $\mathbf{\widetilde{B}}(i)$ of LTI system.
    \EndFor
    \State $\mathbf{\widetilde{W}_A},\mathbf{\widetilde{W}_B} \gets$ Find reduced weights by solving the $r$-rank Procrustes problem for the system:
    \begin{equation}
   \left \{
\begin{aligned} 
     \mathbf{\widetilde{A}}(i) &=\widetilde{\mathbf{W}}_{\mathbf{A}}(\phi(\boldsymbol{\Theta}_i) \otimes \mathbf{I_s})\\
     \mathbf{\widetilde{B}}(i) &=\widetilde{\mathbf{W}}_{\mathbf{B}}(\psi(\boldsymbol{\Theta}_i) \otimes \mathbf{I_{in}})
\end{aligned},
 \right .
\end{equation}
    with every parameter as input and every corresponding LTI weight as output (reduced by the POD transformation).
  \end{algorithmic}
\end{algorithm}

With this procedure, all training is performed in the reduced space, reducing overhead from training.
Both algorithms will be tested to show that both achieve very close performance in terms of one-step prediction, demonstrating that  local identification can be carried out in the reduced-order space.
Since we assumed $n_s$ to be a large number, performing multiple least-squares identification of LTI systems in the case of a full-model LPV identification might be costly.
If ones uses Algorithm \ref{alg:local:LPV:B}, then it is possible to obtain a model by training multiple $r-$sized LTI systems.
Both Algorithm \ref{alg:local:LPV} and \ref{alg:local:LPV:B} solve the $r-$rank Procrustes problem over $r \times r$ and $r \times n_u$ output matrices (same as solving for $r^2$ and $rn_u$ outputs), and $2rn_p$ + $2rn_u$ inputs.
The problem itself is already presented in the latent space, and by performing a $r-$rank Procrustes approximation, one just has to solve an $r-$truncated SVD instead of the full SVD to find the pseudo-inverse, and thus the model.

\section{Applications} \label{sec:application}

This section reports results from applying global and local LPV-system identification to the discretized numerical scheme for solving the partial differential equation (PDE) regarding the diffusion system in a 1-d line.
Such application serves as a proof of concept for the proposed DMD-LPV identification method and its model order reduction properties.

\subsection{Parametric Linear Diffusion} \label{sec:exp1}

Consider the following partial differential equation (PDE):
\begin{equation} \label{eqn:lineardiffusion_pde}
  \frac{\partial T}{\partial t} = k(p) \frac{\partial^2 T}{\partial x^2} - w\frac{\partial T}{\partial x} 
\end{equation}
where $T(x,t)$ is the unknown solution, and $w=0.1$ is the so-called advection velocity, a constant of the system.
The PDE \eqref{eqn:lineardiffusion_pde} is known as the diffusion equation and is widely used in thermodynamics.
In that sense, $T(x,t)$ represents the temperature of a system at a given point in space and time.

In this application, we consider the functions that define the diffusion gain $k(p)$ to be a polynomial on the parameter $p$:
\begin{align}
    k(p) &= 0.1 + 0.05p + 0.01p^2 + 0.03p^3
\end{align}

Further, the spatial domain of the PDE is a straight-line segment. 
  Therefore, the domain is $x \in \Omega$, $\Omega = (0,1)$.
The boundary conditions are as follows:
\begin{equation}
\left \{ \begin{aligned}
    T(0,t) &= u(t)\\
    \partial_x T(1,t) &= 0
\end{aligned}
   \right .
\end{equation}
The boundary $T(0,t)$ is seen as a point where the temperature of the system is controlled to stay at a fixed level $u(t)$, modeled as the sole input to the system.
  On the other hand, the boundary condition $\partial_x T(1,t)$ means that there is no heat leaving the system  at $x = 1$, therefore it finds itself in equilibrium at that given point in space.

PDE systems can be seen as dynamic systems with an infinite number of states.
From the point of view of identification and control, we tend to approximate them into an ODE system \cite{Thomee20011} by discretizing the space domain.
ODEs have a wide array of tools to perform identification and control compared to PDEs.
However, to approximate a PDE into an ODE through space discretization, a large number of points in the discretized grid of the space domain are needed.
Hence, any PDE-described system with a sufficiently fine grid can be labeled as a large-scale dynamic system because of the number of states.

To simplify the discussion, we take the domain defined for the nonlinear one-dimensional diffusion equation and distribute it into several equally spaced points with distance $h$. 
The finite difference method, one such PDE discretization method, states that, through center approximation:
\begin{equation}
\left \{ \begin{aligned}
 \frac{\partial T(x_i,t)}{\partial x_i} &= \frac{(T(x_i + h,t) - T(x_i-h,t))}{2h}\\
  \frac{\partial^{2} T(x_i,t)}{\partial x^2} &= \frac{(T(x_i + h,t) - 2T(x_i,t) + T(x_i - h,t))}{h^2}
\end{aligned}
\right .
\end{equation}
for an arbitrary value $x_i$ inside the domain and the interval $h$.
Also, we consider the following, omitting the time-dependence in the notation, as in state-space notation:
\begin{equation} \label{eqn:state_indexing}
\left \{ \begin{aligned}
  x_0 &= 0\\
   x_{i+1} &= x_i + h\\
  x_{N+1} &= 1\\
  T(x_i,t) &= T_i
\end{aligned} \right . 
\end{equation}
where $x_0$ and $x_N$ are defined by the edge of the original, continuous domain.
The value of $T$ at those points is defined by the boundary conditions, which is why they do not correspond to a state for the ODE.
This means that the number of states on the discretized system is $N = \left \lceil{1/h}\right \rceil - 1$ for $h < 1$ (the actual length of the line segment considered as domain $\Omega$).

Regarding the finite-difference numerical scheme, the diffusion equation becomes as follows, for any point of $x_i$ inside $\Omega$:
\begin{equation} \label{eqn:lineardiffusion_fdm1}
  \dot{T}_i =  k(p)\frac{T_{i+1} - 2T_i + T_{i-1}}{h^2} - w\frac{T_{i+1} - T_{i-1}}{2h} 
\end{equation}

Or, in a vectorized form:
\begin{equation} \label{eqn:lineardiffusion_fdm2}
  \dot{T}_i =
  \left(
   \begin{bmatrix}
     1\\
     -2\\
     1
  \end{bmatrix}k(p)/h^2
      +
 \begin{bmatrix}
     1\\
     0\\
     -1
  \end{bmatrix}(- w/(2h))
  \right)^T
  \begin{bmatrix}
      T_{i+1}\\
      T_i\\
      T_{i-1}
  \end{bmatrix}
\end{equation}

The first boundary condition defines the value that $T_0$ assumes, which is expressed as follows:
\begin{equation}
    T_0 = u(t)
\end{equation}
Therefore:
\begin{equation} \label{eqn:linearfdm_boundary1}
  \dot{T}_1 =
  \left(
  \begin{bmatrix}
     1\\
     0
  \end{bmatrix}(- w/(2h)) +
  \begin{bmatrix}
     1\\
     -2
  \end{bmatrix}k(p)/h^2\right)^T
  \begin{bmatrix}
      T_2\\
      T_1
  \end{bmatrix}\\
  + \left (k(p)/h^2 + w/2h\right )u(t)
\end{equation}

The second boundary condition, a Neumann boundary condition \cite{Thomee20011} in $T_{N+1}$, through the backward discretization method, basically states in this case that $T_{N} = T_{N+1}$ where $T_{N+1}$ corresponds to the point where $x = 1$ and $T_{N}$ is the closest point to it.
The state $T_N$ is described as:
\begin{equation} \label{eqn:linearfdm_boundary2}
  \dot{T}_N =
  \left(\begin{bmatrix}
     1\\
     -1
  \end{bmatrix}\left(- k(p)/h^2-w/(2h)\right)\right)^T
  \begin{bmatrix}
      T_N\\
      T_{N-1}
  \end{bmatrix}
\end{equation}

The following LPV state space then represents the complete system:
\begin{equation}
    \mathbf{\dot{T}} = \left(\mathbf{A_0} +\mathbf{A}(p)\right)\mathbf{T} + \left (\mathbf{B_0} + \mathbf{B}(p)\right)u(t)
\end{equation}
where $\mathbf{T}$ is the vector of states along the $1D$ domain, indexed according to Equation \eqref{eqn:state_indexing}, and:
\begin{footnotesize}
\begin{align}
 \mathbf{A_0} &=
 -(w/2h)\mathbf{D_1}\\
 \mathbf{A}(p) &= (k(p)/h^2)
 \mathbf{D_2}\\
 \mathbf{B_0} &=(w/2h)
 \begin{bmatrix}
     1\\
     0\\
     0\\
     \vdots\\
     0
 \end{bmatrix} \\
 \mathbf{B}(p) &=(1/h^2)
 \begin{bmatrix}
     1\\
     0\\
     0\\
     \vdots\\
     0
 \end{bmatrix}k(p)\\
 \mathbf{D_1} &=
 \begin{bmatrix}
     0 & 1 & 0 & 0 & \hdots & 0\\
     -1 & 0 & 1 & 0 & \hdots & 0\\
     0 & -1 & 0 & 1 & \hdots & 0\\
     \vdots & \vdots & \vdots & \vdots & \ddots & \vdots\\
     0 & 0 & 0 & -1 & \hdots & 1
 \end{bmatrix}\\
 \mathbf{D_2} &=
 \begin{bmatrix}
     -2 & 1 & 0 & 0 & \hdots & 0\\
     1 & - 2 & 1 & 0 & \hdots & 0\\
     0 & 1 & - 2 & 1 & \hdots & 0\\
     \vdots & \vdots & \vdots & \vdots & \ddots & \vdots \\
     0 & 0 & 0 & 0 & 1 & -1
 \end{bmatrix}
\end{align}
\end{footnotesize}

\subsubsection{Implementation, Simulation and Data Gathering}

To simulate the system, we set $h = 0.02$ so that there are $n_s = 49$ states in the discretized system, and the number of inputs is $n_u = 1$.
The discretization is implemented using Python and simulated in time using Runge Kutta (4th order) at $\delta t = 10^{-3}$ so that the dynamics of the system is thoroughly captured.
The sampling time for obtaining a data point was $T_s =  10^{-3}$.

To gather data for the nonlinear diffusion equation system, we consider that $\forall t, u(t) \in [0,4]$ and $p \in [0,1]$.
Then, we excitate the system with an APRBS (Amplitude-modulated pseudo-random binary signal) signal, considering a minimum step size for the control action and parameter $p$ equivalent to $10000$ time steps.
In other words, the system is excited for identification through a stair signal varying randomly between the aforementioned value sets for both the parameters and the control signal.

\begin{figure}[h!]
    \centering
    \includegraphics[width=0.9\linewidth]{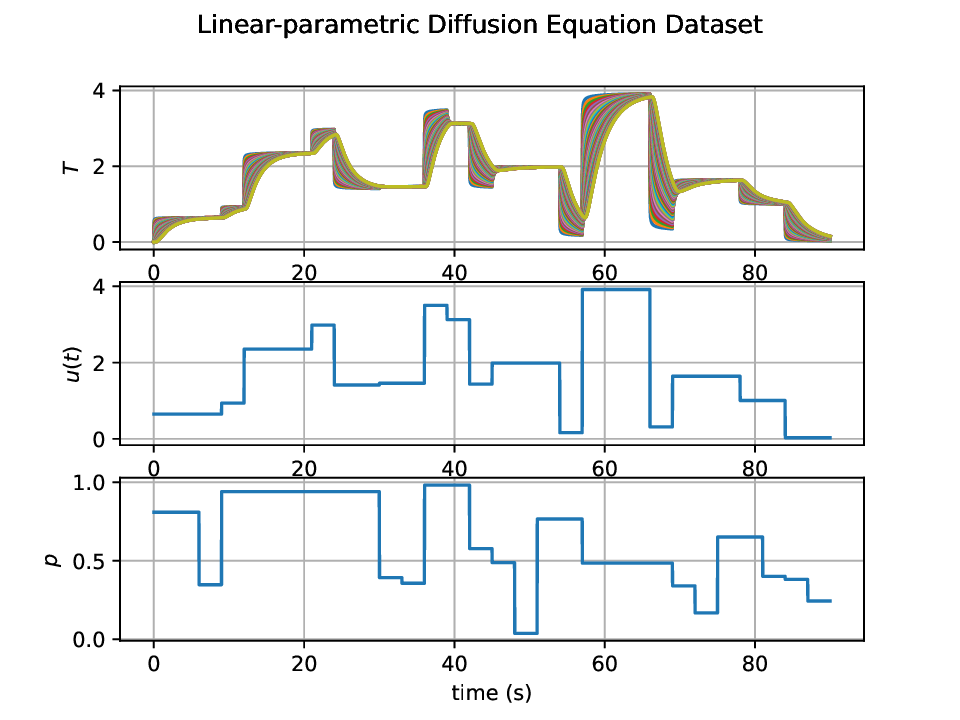}
    \caption{Data-set employed for the linear parametric diffusion equation system, where the top plot is the dynamics of the whole $n_s = 49$ points measured for $T$ (with the closest point to $x=1$ being depicted as yellow), the middle plot is the excitation signal for the sole system input $u(t)$, and the bottom plot shows the values selected for $p$ along the simulation. Note that, besides the plot being in seconds, this figure corresponds to $N = 90000$ data points.}
    \label{fig:lineardiff_equation_dataset}
\end{figure}

Figure \ref{fig:lineardiff_equation_dataset} depicts the resulting simulation, which is used as a training set for all the experiments.

This data-set will be used in the experiment to provide a metric to compare each instance of the DMD-LPV parameters, with different Procrustes and POD ranks.
To simplify the analysis, regularization shall not be considered for this work.
The metric is the mean-squared error (MSE) associated with the obtained model and this data-set, being is defined as follows:
\begin{equation}
    MSE = \frac{1}{Nn_s}\sum_{k=1}^N\sum_{i=1}^{n_s}\left(y_{i,data}[k] - y_{i,model}[k]\right)^2
\end{equation}

\subsubsection{Structure Selection}

Since the main objective of this work is to evaluate aspects of model order reduction tools on LPV and quasi-LPV systems, we chose two structures for the LPV model to be identified:
\begin{itemize}
    \item \textbf{Exact Structure}: The model and the identified system have the same structure. This means that the LPV model considered for identification assumes the following features:
    \begin{equation}
        \boldsymbol{\phi(p)} = \boldsymbol{\psi(p)} = 
        \begin{bmatrix}
            1\\
            p\\
            p^2\\
            p^3
        \end{bmatrix}
    \end{equation}
    In this case, $N_f = 4$.
    \item \textbf{Structural Underestimation}: For this case, we assume that, in selecting the model structure, we miss the cubic parameter-feature relation, therefore:
    \begin{equation}
        \boldsymbol{\phi(p)} = \boldsymbol{\psi(p)} = 
        \begin{bmatrix}
            1\\
            p\\
            p^2
        \end{bmatrix}
    \end{equation}
    For this parameter selection, $N_f = 3$.
    \item \textbf{Structural Overestimation}: Instead of missing one feature, this time, the model design experiments with adding $p^4$ to the parameter feature poll without knowing that it does not exist in the real system. Then, 
    \begin{equation}
        \boldsymbol{\phi(p)} = \boldsymbol{\psi(p)} = 
        \begin{bmatrix}
            1\\
            p\\
            p^2\\
            p^3\\
            p^4
        \end{bmatrix}
    \end{equation}
    For this polynomial, $N_f = 5$.
\end{itemize}

Since this evaluation is a proof of concept, we chose the exact model as a test to evaluate the performance of the DMD-LPV in a scenario where the full-order Least Squares method achieves nearly exact precision. This allows us to assess only the properties related to the reduction performed by DMD.
Additionally, the other two model structures selected are designed to test the model's performance in the presence of small structural model-system uncertainties.

\subsubsection{Procrustes Problem}

We now perform an experiment to assess how close the Procrustes problem \eqref{eqn:procustes_lpv} approaches\footnote{Refer to Section \ref{subsec:Glb-LPV} for details.} the least squares when applied for identification of the linear parametric diffusion equation.
   We evaluate different rank values $r \in \{10,20,30,40,50,60,80,100,120\}$ for the singular values and vector truncation, applying Algorithm \ref{alg:global:LPV} without performing the POD operation. 
Then, we calculate the training MSE error for each solution, while also computing the full least-squares for the system (full-rank Procrustes problem). 
Note that the maximum rank for the Procrustes problem is actually $n_s \times N_f$ ($4 \times 49 = 196$), since it is the number of model features.
Solving a Procrustes problem with such rank is the same as solving the Least Squares Problem.
This number is $N_f$ times larger than the number of states, which is why it is expected that in global LPV identification, the Procrustes rank and the POD rank are not the same.

\begin{table}[]
    \caption{Results of the Procrustes Problem experiment for the linear diffusion equation, measured by the mean squared error over the training data for different ranks for the weight matrices. }
    \label{tab:procrustes_result_linear}
    \centering
    \begin{tabular}{|c|c|c|c|} 
      \cline{2-4}
     \multicolumn{1}{c}{}   & \multicolumn{3}{|c|}{MSE (training)} \\  \hline
      Rank & Exact & Underestimated & Overestimated \\\hline
      $10$ & 1.47e-05 & 6.37e-06 & 3.41e-05  \\ \hline
      $20$ & 6.53e-08 & 1.35e-08 & 1.46e-07 \\ \hline 
      $30$ & 3.30e-10 & 2.94e-11 & 1.74e-09 \\ \hline 
      $40$ & 3.96e-12 & 1.21e-12 & 2.82e-11 \\ \hline 
      $50$ & 5.56e-14 & 9.72e-13& 4.37e-13 \\ \hline
      $60$ & 1.43e-14 &9.06e-13 & 1.76e-14 \\ \hline
      $80$ & 1.32e-14 &7.30e-13 & 2.11e-16 \\ \hline
      $100$ & 1.25e-14 &6.88e-13 & 1.89e-16 \\ \hline
      $120$ & 1.23e-14 & 6.933e-13 & 1.7996e-16 \\ \hline
      Full ($196$)  & 1.23e-14 & 6.943e-13 & 1.7974e-16 \\
      \hline
    \end{tabular}
\end{table}

Table \ref{tab:procrustes_result_linear} depicts the results from the experiment that performs only the Procrustes reduction without performing POD.
The full rank differs for each case as a result of each employed model considering a different polynomial: $147$ for the underestimated (quadratic polynomial) case, $196$ for the exact (third-degree polynomial) case, and $245$ for the overestimated (fourth-degree polynomial) case.
The main idea of solving the Procrustes problem in the context of DMD-LPV is to perform an efficient computation of the weight matrix through rank limitation, which would cost less than solving the full Least Squares problem.
The underestimated and exact cases show diminishing returns (being in the same decade) at rank $50$, while the overestimated case should show such diminishing returns at rank $80$.
Since the very low error associated with the overestimated case is likely a consequence of overfitting, solving the rank $50$ Procrustes problem is as efficient as solving the full least squares problem, even accounting for a lack of precision.

Procrustes rank $40$, $50$, and $60$ make interesting test cases for the next experiment, as $40$ corresponds to before the diminishing return starts, $50$ is where the exact and underestimated cases enter the same decade  in the log scale as the full least squares error, and $60$ has a small MSE improvement over rank $50$.

\subsubsection{POD Reduction}

To evaluate the POD reduction, we applied Algorithm \ref{alg:global:LPV} with the POD operation for ranks $r \in\{1,5,10,15,20,25,30,35,40,45,49\}$ which correspond to the dimension of reduced state variable $\mathbf{z}$.
Since our interest is to find the smallest rank that does not degenerate the system behavior, we tested values for where the diminishing returns on training MSE started to appear.
This is why we selected ranks $\{40,50,60,80\}$ of the Procrustes reduction.
Procrustes Rank of $40$ was right before the MSE diminishing returns started, and $40$, $50$, and $60$ are the earliest stages.

\begin{figure}
    \centering
    \includegraphics[width=0.7\linewidth]{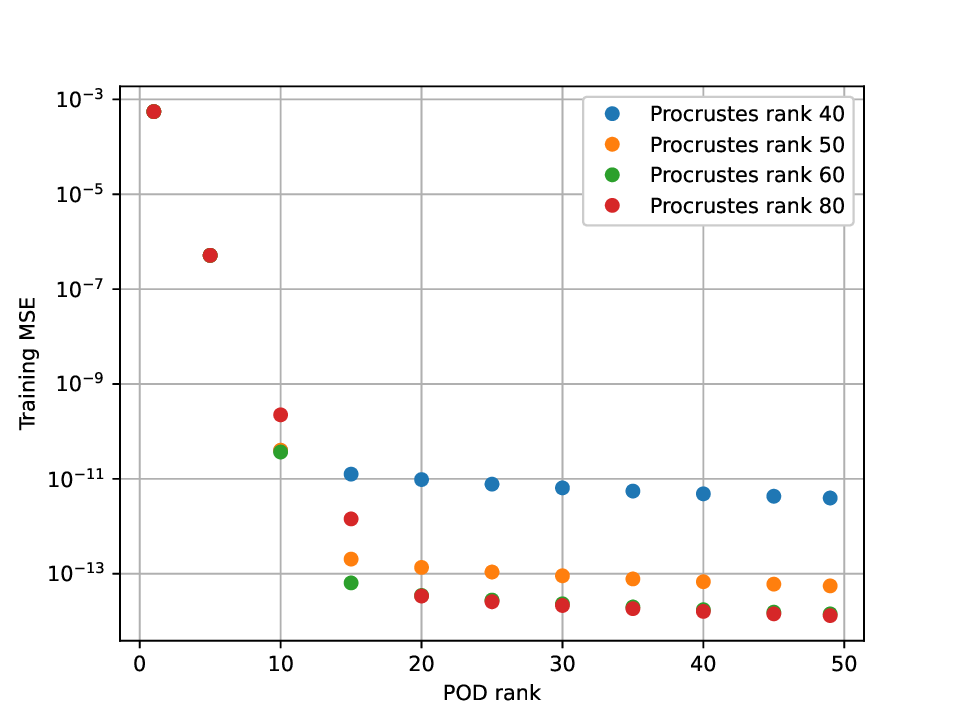}
    \caption{Plot of the training data MSE in log scale as a function of the POD reduction rank for the exact case  ($N_f=4$). Each curve represents a different value for the Procrustes Problem rank.}
    \label{fig:pod_error_plot}
\end{figure}

Figure \ref{fig:pod_error_plot} showcases the result of the experiment in log scale.
It shows that for the Procrustes ranks depicted, there is diminishing returns on POD rank $15$ and $20$, which represent around $31\%$ and $41\%$ of the original number of states considering the exact case ($N_f=4$).

Figure \ref{fig:pod_error_plot} corroborates the fact that the identification error is degenerated by a factor of $100$ between Procrustes ranks $40$ and $50$ (two decades in log scale). Meanwhile, a Procrustes rank of $60$ shows very similar error behavior to a Procrustes rank of $80$ for any rank, even behaving better at a POD rank of $15$.
This is why only Procrustes of rank $50$ and $60$ are considered for experiments on the underestimated and overestimated models.

\begin{figure}
    \centering
    \includegraphics[width=0.7\linewidth]{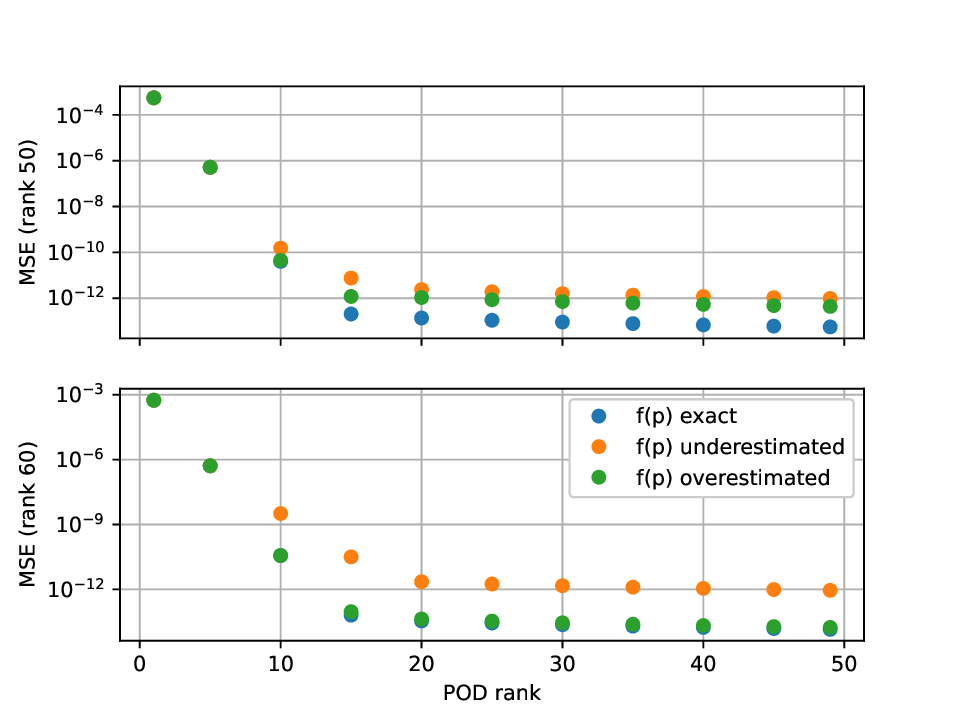}
    \caption{Plot of the training data MSE for Procrustes ranks 50 and 60 in log scale as a function of the POD reduction rank, now further depicting the results for the underestimated and the overestimated case. Each curve represents a different value for the Procrustes Problem rank. The first plot depicts the underestimated, exact, and overestimated model for the rank $50$ case, and the second plot is analog, but for the rank $60$ case. Notice that for the $r = 60$ case, the performance of the overestimated model is close to the exact model.}
    \label{fig:pod_error_underandover_plot}
\end{figure}

Figure \ref{fig:pod_error_underandover_plot} depicts the POD error for Procrustes rank $50$ and $60$, considering the exact, underestimated, and overestimated model.
One factor that must be considered when analyzing errors for the overestimated case is that they come from overfitting the model.
The $p^4$ monomial does not exist in the real system, yet the training error is lower, which does not hold for untrained data.
This is due to the larger sensitivity of the overestimated model.
In fact, the rank $ 50 $ performs some regularization (note that no model regularization is employed), with the error of the overfitted case coherently being larger than the exact case.
The error of the overestimated case is close to the underestimated case for higher POD ranks.
Also, the error incurred in the underestimated case is lower in the rank $50$ case than in the rank $60$ case for lower POD ranks.
The discussion of these results is further elaborated later on with a separate simulation.

\subsubsection{Local Identification}

Local identification is trivial since the parametric linear diffusion model is a true LPV and not a quasi-LPV.
To identify the linear parametric diffusion system, we provide a different cluster of data sets where the system is excited with a fixed $p$ and an APRBS signal for $u$. 
The values of $p$ for this experiment are drawn from the set $\{0,0.1,0.2,0.3,0.4,0.5,0.6,0.7,0.8,0.9,1\}$.
Here, we apply Algorithms \ref{alg:local:LPV} and \ref{alg:local:LPV:B}  to obtain a reduced-order model locally, aiming to showcase that both algorithms achieve a nearly equivalent performance, thus validating Algorithm \ref{alg:local:LPV:B} as a local model order reduction method.
   However, Algorithm \ref{alg:local:LPV:B} is computationally cheaper because the identification of LTI systems is performed in reduced-order space, whereas Algorithm \ref{alg:local:LPV} carries out the LTI identification in the full state space.

    We use the global training data as a metric despite training each LTI model in a $12000$ timestep simulation.
This means that we are intrinsically performing a one-step prediction test. Therefore, the MSE value is supposed to be higher than an error measure over training data.
In fact, the full-order local identification with the exact polynomial performs with an MSE of $1.39 \times 10^{-7}$.

\begin{table}[htb]
    \centering
        \caption{MSE Result from the global case training data-set for the local identification, identifying full order LTI systems. The rank in the first column is both the POD rank and the Procrustes rank to perform the final weight identification, which are the same. The $\Delta$ means that the value depicted is the difference between the MSE in question and the Exact MSE for that rank value. A negative value is a lower error, and a positive value is a larger error.
        This choice of depiction is because the difference in errors is quite small.}     \label{tab:local_result_linear1}
    \begin{tabular}{|c|c|c|c|}
       \cline{2-4}
     \multicolumn{1}{c}{}   & \multicolumn{3}{|c|}{MSE (one-step prediction of global training data-set)} \\\hline
      Rank & Exact & Underestimated ($\Delta$) & Overestimated ($\Delta$) \\\hline
      $1$ & $2.91\times 10^{-4} $& $2\times 10^{-9}$ & $2\times 10^{-9}$ \\ \hline
      $5$ & $1.26\times 10^{-6}$ & $4\times 10^{-10}$ & $4\times 10^{-10}$ \\ \hline 
      $10$ & $6.82\times 10^{-7}$ & $1.7\times 10^{-10} $ & $1.7\times 10^{-10}$ \\ \hline 
      $15$ & $4.55\times 10^{-7}$ & $1.7\times 10^{-10} $ & $1.7\times 10^{-10}$\\ \hline 
      $20$ & $3.41\times 10^{-7}$ &$1.3\times 10^{-10}$& $1.3\times 10^{-10}$\\ \hline
      $25$ & $2.73\times 10^{-7}$ &$1.1\times 10^{-10}$& $1.1\times 10^{-10}$ \\ \hline
      $30$ & $2.28\times 10^{-7}$ &$9\times 10^{-11}$ & $9\times 10^{-11}$\\ \hline
      $35$ & $1.95\times 10^{-7}$ &$7\times 10^{-11}$ & $7\times 10^{-11}$\\ \hline
      $40$ & $1.71\times 10^{-7}$ & $7\times 10^{-11}$ &$7\times 10^{-11}$\\ \hline
      $45$  & $1.52\times 10^{-7} $& $6\times 10^{-11}$& $6\times 10^{-11}$ \\ 
      \hline
    \end{tabular}
\end{table}

Table \ref{tab:local_result_linear1} shows the results for the local LPV identification experiment with the full-order LTI training.
The MSE was measured for the global training set, considering both the exact, underestimated, and overestimated models for the different rank values depicted.
The local identification allows the Procrustes and POD reduction at the same rank since the maximum rank for the Procrustes problem is $N_f \times r_{pod}$.

This method looks robust regarding polynomial structural uncertainty for one-step prediction, since the MSE error difference for the three cases is relatively small.
The reduced-order model reaches the same MSE decade as the full order model at a rank as small as $10$, which means that performing the reduced-order identification with such rank would provide a close representation of the whole system.
We also considered ranks $5$ and $15$ for the results showcased in Section \ref{sec:results_showcase} to assess the model performance in a full simulation problem before and after the diminishing returns.

\begin{table}[htb]
    \caption{MSE result from the global case training data-set for the local identification, identifying LTI systems in the latent space defined by the POD transformation. The rank in the first column is both the POD rank and the Procrustes rank to perform the final weight identification, which are the same. The $\Delta$ means that the value depicted is the difference between the MSE in question and the Exact MSE for that rank value. A negative value is a lower error, and a positive value is a larger error.
        This choice of depiction is because the difference in errors is quite small.}
    \label{tab:local_result_linear2}

    \centering
    \begin{tabular}{|c|c|c|c|}
     \cline{2-4}
     \multicolumn{1}{c}{}   & \multicolumn{3}{|c|}{MSE (one-step prediction of global training data-set)} \\\hline
      Rank & Exact & Underestimated ($\Delta$) & Overestimated ($\Delta$)\\\hline
      $1$ & $4.99\times 10^{-6}$& $3\times 10^{-18}$ & $3\times 10^{-18}$ \\ \hline
      $5$ & $1.67\times 10^{-6}$ & $1.2\times 10^{-19} $& $1.3\times 10^{-19} $ \\ \hline 
      $10$ & $6.82\times 10^{-7}$ & $-1\times 10^{-19}$  & $-2\times 10^{-19}$\\ \hline 
      $15$ &$ 4.55\times 10^{-7}$ & $-2\times 10^{-19}$ & $-2\times 10^{-19}$ \\ \hline 
      $20$ & $3.41\times 10^{-7}$ & $3\times 10^{-17}$ & $3\times 10^{-17}$ \\ \hline
      $25$ & $2.73\times 10^{-7}$ &$-1\times 10^{-14}$& $-1\times 10^{-14}$ \\ \hline
      $30$ & $2.28 \times 10^{-7}$ &$-1\times 10^{-13}$ & $-1\times 10^{-13}$\\ \hline
      $35$ & $1.95\times 10^{-7} $&$3\times 10^{-11}$ & $ 6\times 10^{-11} $\\ \hline
      $40$ & $1.71\times 10^{-7}$ & -$1\times 10^{-11}$ & -$1\times 10^{-11}$\\ \hline
      $45$  & $1.52\times 10^{-7}$ & -$1\times 10^{-11}$ & -$1\times 10^{-11}$ \\
      \hline
    \end{tabular}
\end{table}

Table \ref{tab:local_result_linear2} showcases the same experiment as the one depicted in Table \ref{tab:local_result_linear1}, however, with the alternate proposed local LPV identification method.
The purpose of this experiment was to evaluate whether using the POD transformation corresponding to the full LPV model would serve as a valid linear transformation to identify the local LTI systems in a reduced latent space, and then employ such a transformation to map the reduced LPV model states back into full order.
It turned out that both methods showed similar efficacy in terms of the one-step prediction problem, which validates the use of the POD transformation in local LPV identification.
Notice that the error difference ($\Delta$) between the model mismatch cases and the exact case might even miss machine precision for the lower rank cases, which are the ones of interest in this application.

Overall, these experiments show that a model order reduction in a local linear environment might be more computationally efficient than a global one, while using the same decade criterion as the full-order error criterion.

\subsubsection{Results and Discussion} \label{sec:results_showcase}

The previous experiments tested different rank (Procrustes and POD) values using the MSE metric for a given data set to train the DMD-LPV model with a global training procedure.
While the local trained DMD-LPV demanded a different kind of data set by design, the global data set was used to test performance in one-step prediction type of problems. 
The previous experiments aimed to find rank values for which the MSE presented diminishing returns considering: the global Procrustes reduction rank, global POD rank, and local POD/Procrustes rank. 
In this section, we perform a simulation experiment considering a completely separate data set. The LPV model only has access to the initial condition. Thus, the runtime is susceptible to cumulative error in contrast to the one-step prediction, where the previous step is given by the data.
The experiment in question considers the following models:
\begin{itemize}
    \item Globally identified DMD-LPV model with (Cartesian product):
    \begin{itemize}
        \item Procrustes Rank: $40$, $50$, and $60$.
        \item POD Rank: $5$, $10$, and $15$.
    \end{itemize}
    \item Locally identified DMD-LPV model rank: $5$, $10$, and $15$.
\end{itemize}

\begin{figure}
    \centering
    \includegraphics[width=0.7\linewidth]{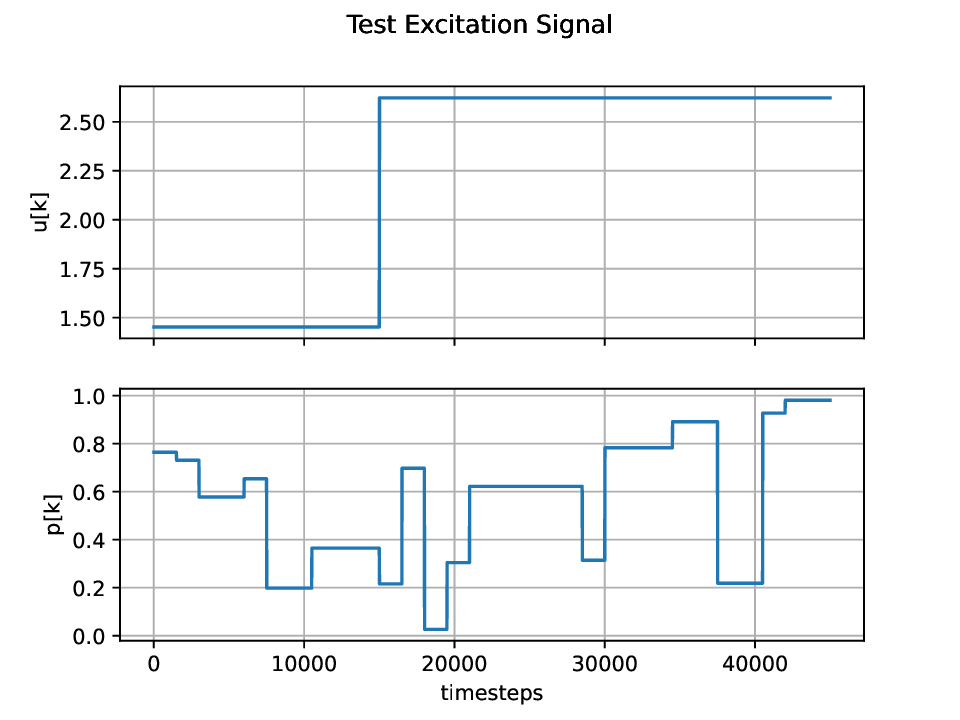}
    \caption{Plot of the excitation signal for both $p[k]$ and $u[k]$ utilized for the simulation experiments.}
    \label{fig:test_excitation}
\end{figure}

Figure \ref{fig:test_excitation} depicts the experiment's excitation signal concerning the parameter $p$ and input $u$.
Although every model performs well in a one-step prediction test similar to the one performed with the local identification set, there are identification models prone to numerical instability for simulation---the model only receives the initial condition and does not use further information from data.

To showcase the simulation result, we consider the signal for $T$ where $x = 0.98$, as it is the furthest point away from the region the input affects $x = 0$, hence that highest order input-to-output dynamic.

\begin{figure}
    \centering
    \includegraphics[width=0.7\linewidth]{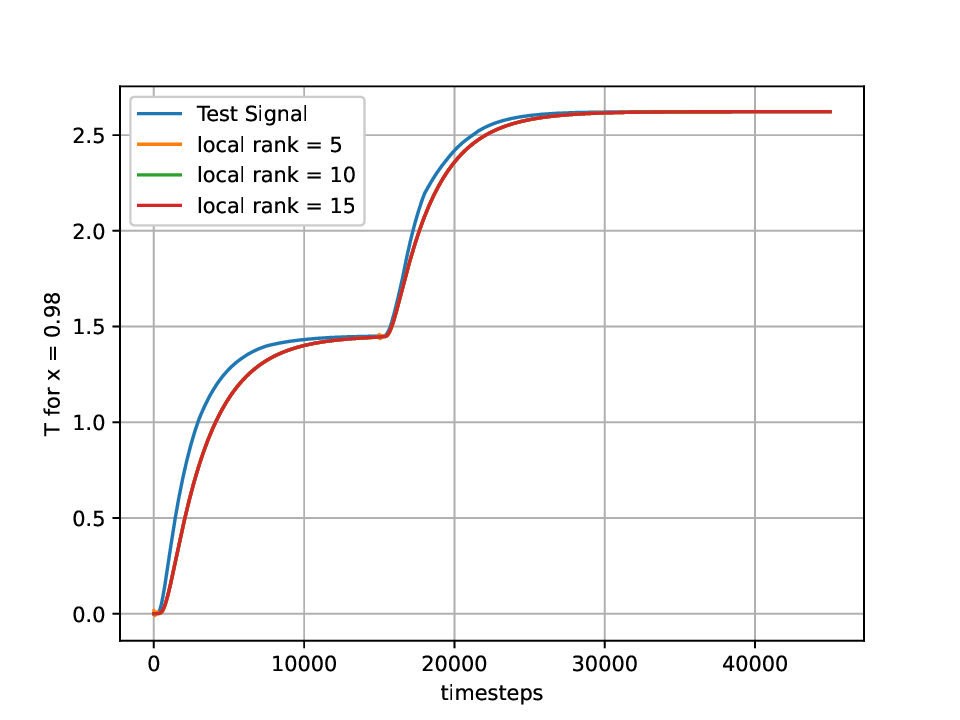}
    \caption{Simulation results for the local, exact polynomial case at the $T$ for the point closest to $x = 1$. The results show that the LPV systems identified locally behaved similarly, showing that a reduction to a dimension of $5$ suffices for model simulation of this application.}
    \label{fig:local_test}
\end{figure}

Figure \ref{fig:local_test} showcases the simulation of the obtained local models in comparison to the test output.
The local models seem less exact than the global models; however, they do not bring issues regarding numerical instability and are also theoretically easier to identify (all models were identified using the reduced LTI procedure).
The three cases tested behaved similarly, which means that a reduction to a dimension of $5$ is perfectly safe to obtain a model simulation of this application.

\begin{figure}
    \centering
    \includegraphics[width=0.7\linewidth]{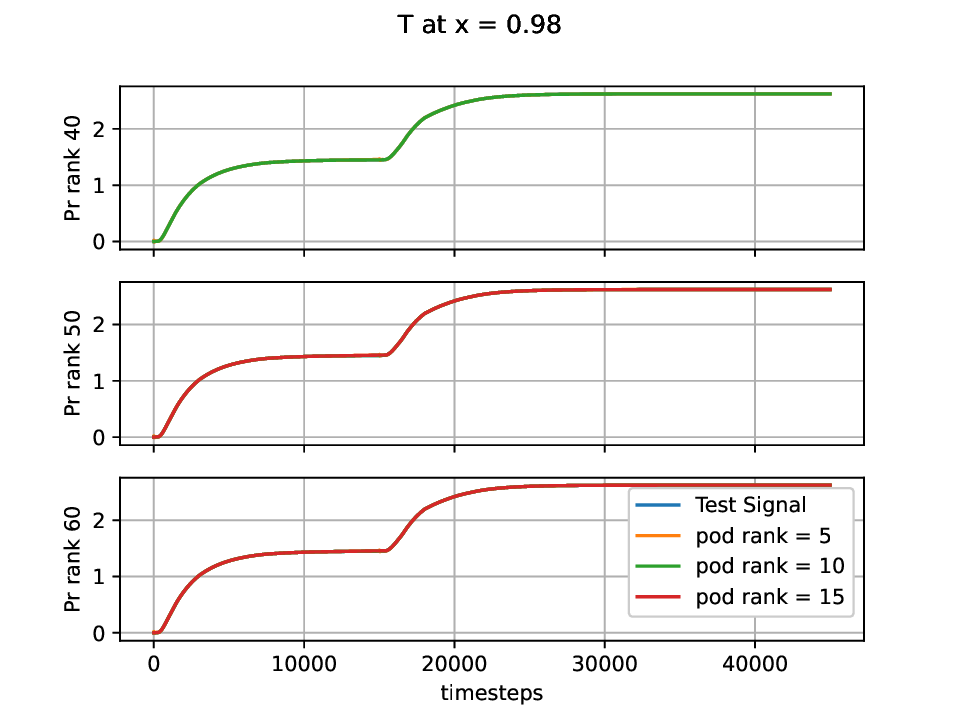}
    \caption{Simulation results for the global, exact polynomial case. Each plot corresponds to a different Procrustes rank. These results show that the LPV system identified globally achieved nearly perfect simulation, considering a Procrustes rank of $40$, $50$, and $60$ in combination with a POD rank of $5$, $10$, and $15$.}
    \label{fig:global_test}
\end{figure}

Figure \ref{fig:global_test} showcases the simulation of the globally obtained models compared to the test output.
Procrustes rank $40$ and POD rank $15$ are not shown because the simulation was halted in the test set because it was numerically unstable.
Testing with other ranks showed that a higher POD rank and a lower Procrustes rank make the identified model prone to numerical instability.
For instance, we did not manage to find a good run for Procrustes rank of $20$.

Overall, global identification might seem superior in performance to local identification, but the proposed local identification procedure is more efficient for finding a reduced-order model that is numerically stable and easier to compute than the full model. 
This might indicate a pattern in which non-intrusive lower-order identification has properties akin to Tikhonov regularization in relation to numerical instability protection.

\begin{figure}
    \centering
    \includegraphics[width=0.7\linewidth]{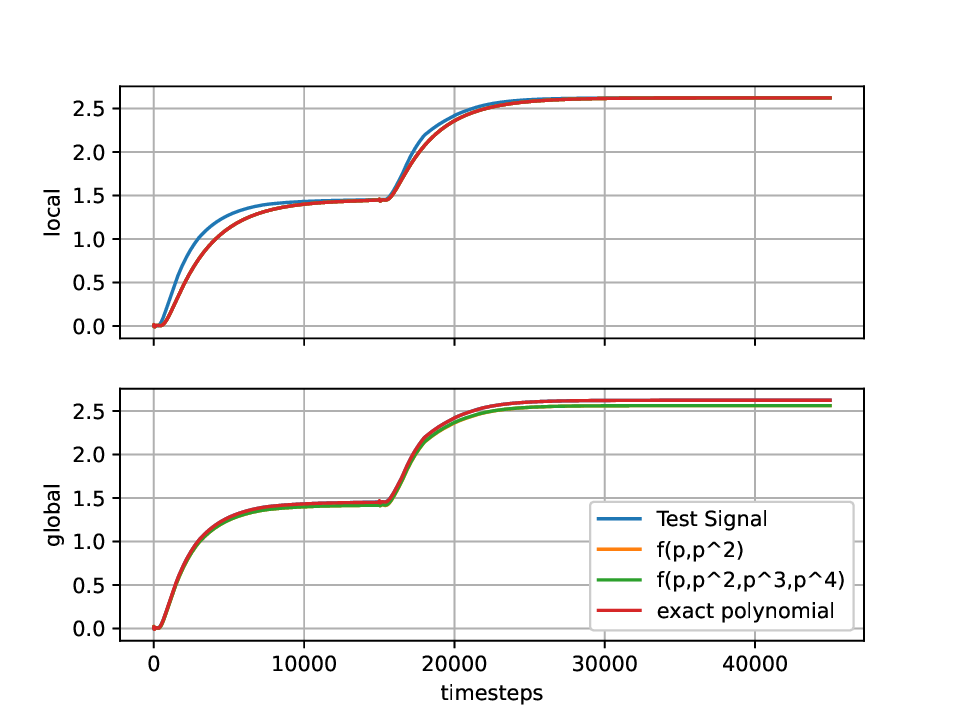}
    \caption{Simulation results for the lowest rank global and locally identified models in terms of underestimated and overestimated polynomial estimation.}
    \label{fig:underover_sim}
\end{figure}

Figure \ref{fig:underover_sim} depicts a simulation involving the DMD-LPV models identified with a structural error, and all simulation plots in this figure refer to the lowest (both Procrustes and POD) ranks for the evaluated models.
The local models seem insensitive to the structural errors committed, corroborated by the one-step prediction experiments.
Meanwhile, the global identification seems more prone to overfitting, as the overestimated model shows steady-state error.

\begin{figure}
    \centering
    \includegraphics[width=0.7\linewidth]{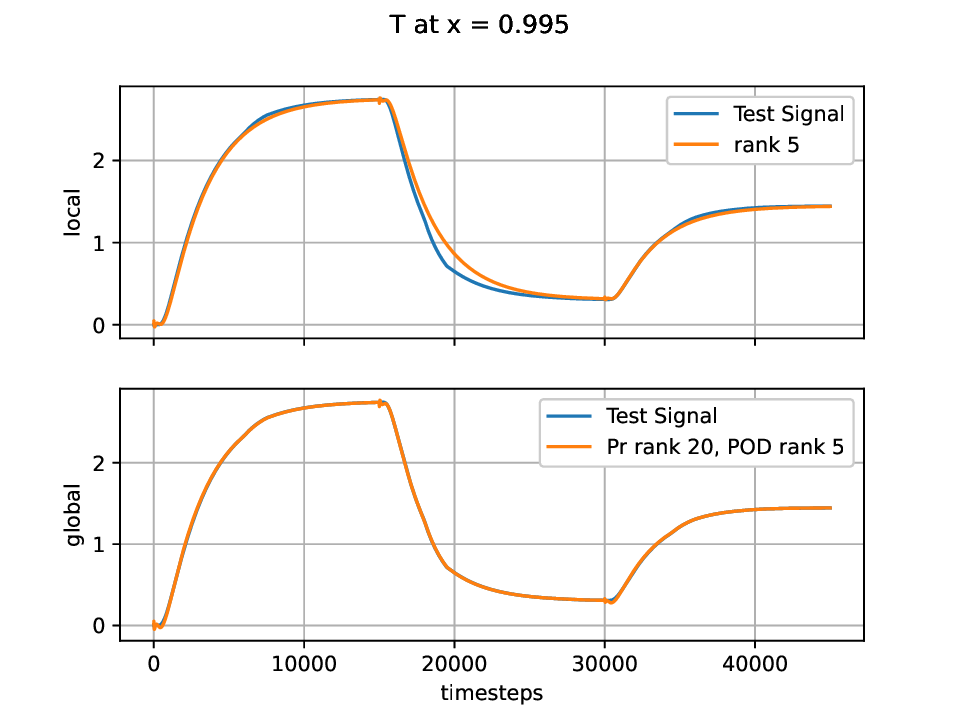}
    \caption{Simulation of identified DMD-LPV models for a version of the linear parametric diffusion equation with $h = 0.005\, (N = 199)$. The rank parameters were selected so that they are the minimum possible to perform a coherent identification.}
    \label{fig:200statediffusion}
\end{figure}

Finally, to better state the state compression power of the DMD-LPV identification, we perform the same described identification procedure in a counterpart of the diffusion equation system with an $h$ of $0.005$ instead of $0.02$.
This entails that the system has a total of $199$ states instead of the $49$ states utilized in the other experiments.
Only a local training procedure for rank $5$ and a global training considering the Procrustes rank $20$ and the POD rank $5$ were performed.
The result, in Figure \ref{fig:200statediffusion}, shows that the simulation of a $200$ state LPV system can be compressed to a $5$ state reduced-order model.
Note that, by the same logic as the rest of this section, the last point in the bar is evaluated; therefore, it presents a $199$ order dynamic in relation to the input, which is the temperature located at $x = 0$, propagating up until $x = 0.995$.

\subsection{Parametric Diffusion Equation with Non-polynomial Gain}

The previous experiments demonstrated a proof of concept for our proposed method by utilizing a simple polynomial function.
This section presents an experiment where the DMD-LPV method is employed in a system with a more complex scheduling function, where:
\begin{itemize}
    \item Two parameters are considered instead of just one.
\item The parametric differential equation features a rational gain function, $k(p_1,p_2)$, instead of the polynomial one in Section \ref{sec:exp1}, thus indicating a more significant mismatch between system and model.
\end{itemize}

Consider the same parametric linear diffusion system defined in Eq. \eqref{eqn:lineardiffusion_pde} but, instead of the polynomial $k(p)$ used throughout the previous experiments, here the function $k$ is given by:
\begin{equation} 
    k(p_1,p_2) = \frac{p_1p_2}{(p_1 + p_2 + 1)^2} \label{eq:sched-f:non-poly}
\end{equation}

Every other definition for the system follows exactly as in Section \ref{sec:exp1}.
We choose the space discretization distance as $h=0.01$ for this experiment, resulting in a system $99$ states consisting of temperature values along a $1d$ grid.
The sampling time for each time step is $T_s = 0.01$ s.

\subsubsection{Structure Selection}

   Regarding the scheduling function for the model trained with DMD-LPV, each respective element is defined as follows: 
\begin{align}
    &\boldsymbol{\phi}_i(p_1,p_2) = \boldsymbol{\psi}_i(p_1,p_2) = p_1^{j_{i,1}}p_2^{j_{i,2}},\forall i \in \{1,2,3,\dots,N_{pol}\}
\end{align}
which corresponds to monomials of the polynomial employed by DMD-LPV, with $j_{i,1}$ and $j_{i,2}$ having arbitrary values for each $i$.
Together, the monomials form an $N_{deg}$ degree polynomial, and the total number of monomials is $N_{pol} = \frac{(N_{deg} + 2)!}{2(N_{deg}!)}$.
Polynomial functions are actually a special case of rational functions, therefore some underfitting (for low $N_{deg}$) and overfitting (for high $N_{deg}$) are expected.

We managed to train this function with data from the $k(p_1,p_2)$ defined in \eqref{eq:sched-f:non-poly}, achieving a relative error of $5.786 \%$ using a polynomial of degree $N_{deg} = 5$ and a parameter interval from $0$ to $1$.
This means that the scheduling function for DMD-LPV applied to this model has $(5 + 2)!/2(5!) = 22$ features from employing the monomials of a $5th$ degree polynomial as a basis function and 99 states, implying that the model itself has $22 \times 99 = 2178$ inputs related to the state, and $22$ inputs pertaining to the control action, which is significantly higher than the number of states.

\subsubsection{DMD-LPV Identification, Results, and Discussion}

Two approaches can be considered for addressing the identification problem: local and global methods.

For this application, the global method is more data-efficient, which is why we chose to implement it for the DMD-LPV approach. Given that the system is stiff, a large number of data points are necessary to accurately capture the behavior of the diffusion equation. This process would need to be repeated for each frozen LTI system in the local approach.

The global method allows for a more straightforward representation of the system dynamics. It enables the full-order model to achieve better performance in representing the system with less data.
However, minimizing the overhead associated with local identification, in terms of the required volume of data, remains an interesting area for future exploration.

For the global identification, we set a Procrustes rank of $110$, which is roughly $5\%$ of the total number of inputs in the DMD-LPV model.
The POD rank was set to $5$, which means roughly $5\%$ of the states.
The control action excitation signal was generated with an APRBS drawn from the uniform distribution, where $u$ falls within the range $[0,4]$ with a minimum step size of $2000$ time steps. For both parameters, an APRBS signal was drawn from a uniform distribution, with $p_1,p_2$ in the range $(0,1]$, having a minimum step size of $150$.
The training gathered $240000$ data points.
The training error was calculated, yielding an error of $6.742\times 10^{-6}$. In contrast, the full least squares method achieved an error of $1.9893 \times 10^{-8}$.
Both models considered  a regularization parameter of $\lambda = 0.05$.

\begin{figure}
    \centering
    \includegraphics[width=0.9\linewidth]{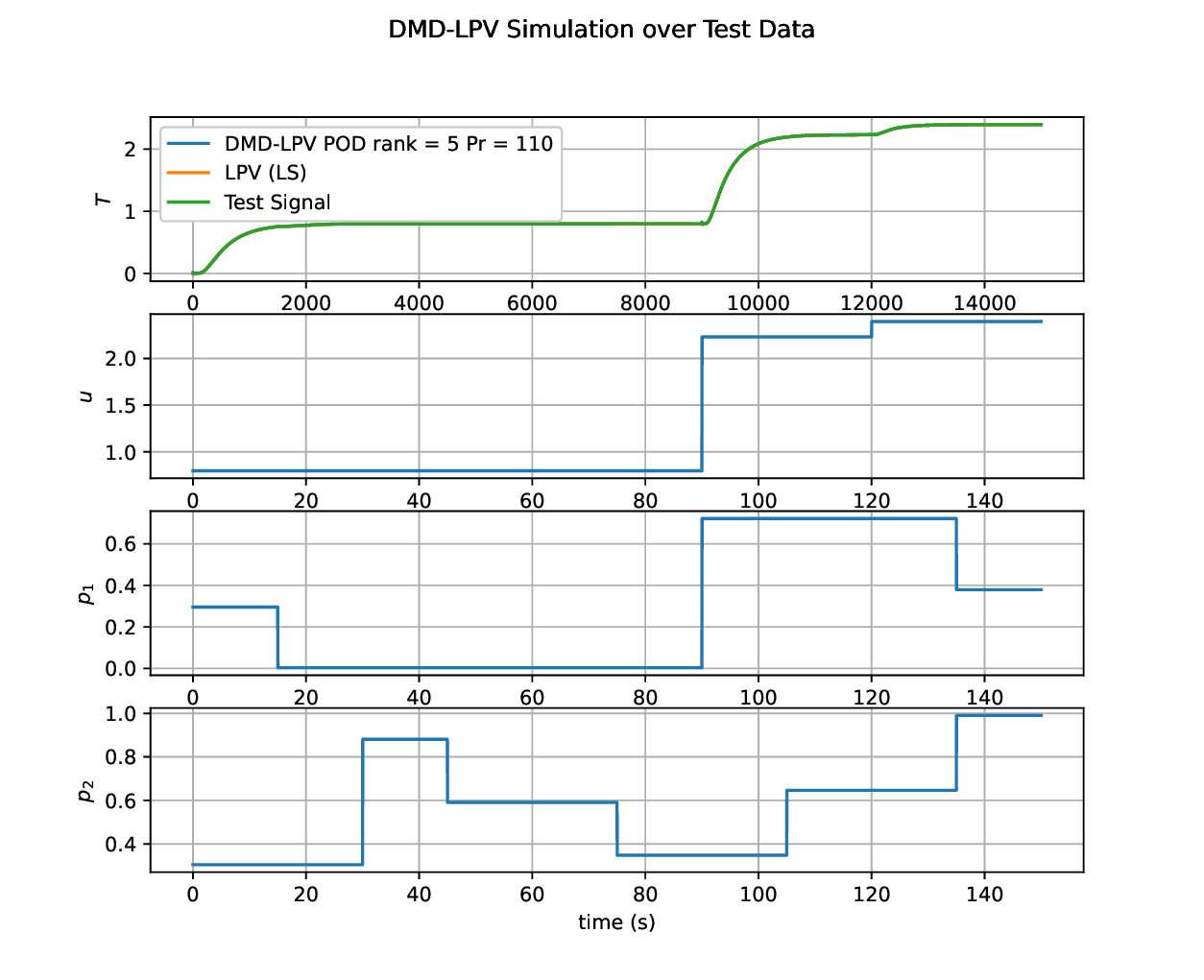}
    \caption{Simulation of the identified DMD-LPV model using a 5th-degree polynomial as the scheduling function for the  diffusion equation with a rational scheduling function over the test data. The green line in the first is the test signal, while the blue line is the DMD-LPV and the orange line is the least-squares trained LPV. The tree lines overlap each other. For comparison, the same LPV model trained at full-order by Least Squares is also plotted as a baseline. The temperature curve is displayed at $x = 0.99$, consistent with other simulation experiments. The rank parameters were chosen to be the minimum required to closely approximate the full-order model.}
    \label{fig:elm_global}
\end{figure}

Figure \ref{fig:elm_global} depicts the test data employed, including the excitation signals for the input and parameters. It also illustrates the simulation of the DMD-LPV model with the $5th$ degree polynomial as the scheduling function, alongside an LPV model trained with Least Squares.

As with the polynomial case, probing the temperature at the other end of the bar from where the manipulated variable is located comes from the reasoning that it is the highest-order input-to-output dynamic in the system.
The performance of the DMD-LPV and the Least Squares model are close for that temperature, even though the DMD-LPV model has only $5$ states (and solves a smaller rank $110$ Procrustes problem), whereas the full LPV model considers $99$ states.
The goal of the proposed method is to obtain such compact LPV models.

Unlike the other case study, there is a structural mismatch between the system and the model, which accounts for the slight differences between the trainable models and the system. 
As in simulation experiments, error propagation is always present, even though the full least squares method and the DMD-LPV approach performed well on the training set.
We also placed the trainable models under the same one-step prediction tests that the experiments for the polynomial case underwent. In these tests, the DMD-LPV achieved a mean squared error of $5.874 \times 10^{-7}$, while the regularized least squares performed with an MSE of $8.069 \times 10^{-10}$ on the same dataset.
The simulation performance was very similar, as the line for the test simulation and the models were similar, even though both models were three decades in the log scale apart in terms of one-step prediction.

\color{black}

\section{Conclusion} \label{sec:conclusion}

We formulated and developed a method to identify a non-intrusive reduced order model of LPV systems based on the DMDc framework.
The so-called DMD-LPV was evaluated for a discretized linear diffusion equation with its diffusion gain defined by a polynomial over a parameter as a proof of concept type of case study, where the developed methodology was put to the test.
We succeeded in performing both global and local identification for such a model, assessing the properties involving the reduction.

Future work will involve employing the DMD-LPV in more complex real-world applications.
Also, suppose we adapt the local methodology to consider quasi-LPV systems. In that case, we can obtain a general identification model for nonlinear systems if the model structure is known.
In the future, working with LPV models in linear fractional transform (LFT) form for identification is a possibility.
Another prospect for future work is using black-box identification functions as the parameter function, such as the Radial Basis Function (RBF) method or kernels.
Since such methods involve a large library of nonlinear functions, the sheer number of features presents an interesting problem for the DMD-LPV to tackle.
Also, the DMD-LPV method was applied to the diffusion equation over a one-dimensional grid.
Theoretically, this algorithm can be extended to higher-dimensional systems, as it does not impose structural assumptions that limit it to only one-dimensional models. 
However, moving to higher-dimensional grids results in a larger number of interconnected states. Therefore, further tests on systems considering higher dimensional grids are interesting for future experimentation.

\section*{Acknowledgments}
This work was funded in part by FAPESC (grant 2021TR2265), CAPES  (grant 88882.182533/2011-01), and CNPq (grant 308624/2021-1).



\bibliographystyle{elsarticle-num-names} 
\bibliography{references}


\end{document}